\documentclass[12pt,preprint]{aastex}

\newcommand{\xmm}{{\it XMM-Newton}\ }

\newcommand{\chandra}{{\it Chandra}\ }

\newcommand{\suzaku}{{\it Suzaku}\ }

\newcommand{\sigadd}{$\sigma_{\rm add}$}
\newcommand{\sigsd}{$\sigma_{\rm sd}$}
\newcommand{\sigsdall}{$\sigma_{\rm sd,all}$}
\newcommand{\kms}{km s$^{-1}$}
\newcommand{\astroh}{ASTRO-H\ }
\newcommand{\vturb}{$v_{\rm turb}$}

\slugcomment{Accepted to ApJ,  \today}

\shorttitle{Perseus cluster}
\shortauthors{Tamura et al.}

\begin{document}

\title{GAS BULK MOTION IN THE PERSEUS CLUSTER MEASURED WITH \suzaku }

\author{T. Tamura, N. Yamasaki, R. Iizuka}
\affil{Institute of Space and Astronautical Science,
Japan Aerospace Exploration Agency,\\
3-1-1 Yoshinodai, Chuo-ku, Sagamihara, Kanagawa 229-8510, Japan
}
\author{Y. Fukazawa}
\affil{Department of Physical Science, Hiroshima University, 1-3-1 Kagamiyama, Higashi-Hiroshima, Hiroshima 739-8526, Japan}

\author{K. Hayashida, S. Ueda}
\affil{
Department of Earth and Space Science, Graduate School of Science, Osaka University, Toyonaka 560-0043, Osaka, Japan}

\author{K. Matsushita, K. Sato}
\affil{Department of Physics, Tokyo University of Science, 1-3 Kagurazaka, Shinjyuku-ku, Tokyo 162-8601, Japan}

\author{K. Nakazawa}
\affil{Department of Physics, The University of Tokyo, 7-3-1 Hongo, Bunkyo-ku, Tokyo 113-0033, Japan}

\author{N. Ota}
\affil{Department of Physics, Nara Women's University, Kitauoyanishimachi, Nara, 630-8506, Japan}

\author{M. Takizawa}
\author{Department of Physics, Yamagata University, 1-4-12 Kojirakawa-machi, Yamagata, 990-8560, Japan}

\begin{abstract}
 We present the results from \suzaku observations of the Perseus galaxy cluster, 
which is relatively close, the brightest in the X-ray sky and a relaxed object with a cool core.
A number of exposures of central regions 
and offset pointing with the X-ray Imaging Spectrometer 
cover a region within  radii of $20'-30'$.
The central data are used to evaluate the instrumental energy-scale calibration with 
accuracy confirmed to within around 300 \kms \, 
by the spatial and temporal variation of the instruments.
These deep and well-calibrated data are used
to measure X-ray redshifts of the intracluster medium.
 A hint of gas bulk motion, 
with radial velocity of about $-(150-300)$ \kms \, 
relative to the main system
was found at $2-4$ arcmin (45--90~kpc) west of the cluster center, 
where an X-ray excess and a cold front were found previously.
 No other velocity structure was discovered.
Over spatial scales of 50--100~kpc 
and within 200~kpc radii of the center, 
the gas-radial-velocity variation is below 300 \kms, 
while over scales of 400~kpc within 600~kpc radii,
the variation is below 600 \kms.
These X-ray redshift distributions
are compared spatially with those of optical member galaxies
for the first time in galaxy clusters.
 Based on X-ray line widths gas turbulent velocities within these regions
are also constrained within 1000--3000 \kms.
These results
of gas dynamics in the core and larger scales
in association with cluster merger activities are discussed
 and future potential of high-energy resolution spectroscopy with \astroh
is considered.

\end{abstract}


\section{INTRODUCTION}
\label{intro}
 
Galaxy clusters are the largest and youngest gravitationally-bound cosmic structure
and form hierarchically through collisions and mergers of smaller systems.
Spatial and radial velocity distributions
of member galaxies and X-ray observations of the intracluster medium (ICM)
have revealed that some systems are still forming and unrelaxed.
Along with gravitational lensing observations, 
these studies have provided dynamical measurements of dark matter in clusters.

As the structure forms, 
gravitational potential governed by dark matter  pulls and thus heats the ICM through shocks.
Sharp X-ray images obtained by \chandra revealed shocks and density discontinuities ( so-called ``cold fronts'') 
suggesting supersonic or transonic gas motions, respectively \citep[see][for a review]{mv07}.
These heating processes would then develop gas turbulence, and accelerate particles, which in turn generate diffuse radio halos and relics.
The activities of central massive black holes 
also disturb and warm up the gas around cluster cores.
The motion of gas should be measured
to facilitate direct understanding of these energy flows.
Constraining total energy distribution in a system
allows the total gravitational mass to be measured, 
as required for precise cosmology.
Energies in gas random or ordered motions 
 could be the key uncertainty to calibrate total mass distribution, 
as suggested by numerical simulations 
\citep[e.g.][]{evrard96, Nagai07, Takizawa10, Suto13}.

Gas bulk or turbulent motions can be measured directly using the Doppler shift or by broadening of X-ray line emissions.
However, 
the limited energy resolutions of current X-ray instruments continues to hinder
such measurements.
For example typical CCD energy resolution is about 130~eV (in full width half maximum; FWHM), 
compared with a possible energy shift of $\sim 20$~eV corresponding to a radial velocity of 1000 \kms,
all at the Fe-K line energy.

The \suzaku XIS \citep{koyama07} is currently the optimal X-ray spectrometer for gas motion search with Fe-K lines, 
as was demonstrated when gas bulk motion was detected in a merging cluster, Abell~2256, by
\cite{tamura2011}.
In addition, XIS data imposed tight constraints on gas motion in a number of clusters, 
as summarized in Table~\ref{tbl:suzaku-res}.

\begin{deluxetable}{lll}
\tablecaption{
Suzaku results of gas bulk motions in clusters.
\label{tbl:suzaku-res}
}
\tablewidth{0pt}
\tablehead{
\colhead{target} & \colhead{result} & \colhead{Reference} 
}
\startdata
Centaurus & $\Delta v < 1400$ km s$^{-1}$  &  1  \\
Oph.   & $\Delta v < 3000 $ km s$^{-1}$ &  2  \\
AMW~7  & $\Delta v < 2000 $ km s$^{-1}$ & 3 \\
A~2319 & $\Delta v < 2000$ km s$^{-1}$ & 4 \\
A~2256 & $\Delta v =  1500 \pm 300 ({\rm stat.}) \pm 300 ({\rm sys.}) $km s$^{-1}$ & 5 \\
Coma  & $\Delta v < 2000$ km s$^{-1}$ & 6 \\
A~3627 & $\Delta v < 800$ km s$^{-1}$ & 7 \\
\enddata
\tablecomments{
Unless stated otherwise, 
values are 90\% limit of gas velocity variation among regions within the cluster.
}
\tablerefs{(1) \cite{ota07}; 
Velocity variation among $2'\times2'$ regions.
(2) \cite{fujita08}.
(3) \cite{2008PASJ...60S.333S}; a hint of a velocity difference between two regions is found.
(4) \cite{sugawara09}.
(5) \cite{tamura2011}; errors are 68\% confidence limit. stat. and sys. mean statistical and systematic errors.
(6) \cite{sato2011}.
(7) \cite{nishino12};  relative velocity of a sub component.
}
\end{deluxetable}

Numerical studies have striven to clarify cluster formation histories 
and the associated origins, development, and dissipation of various types of gas motions.
However, by nature cluster evolution is governed by random processes,
which means systematic measurements of systems in a wide range of evolution stages are 
also required as well as these theoretical efforts.
The geometrically-complex nature of
gas dynamics should be spatially resolved.
For example, gas dynamics in apparently relaxed clusters should be studied
as demonstrated by \cite[][hereafter OTA07]{ota07} using early \suzaku observations of the Centaurus cluster.
To improve the limit of gas bulk motion in a relaxed system,
we analyzed a large set of \suzaku data of the Perseus cluster.

The Perseus cluster, the brightest in X-ray,
is a prototype of a nearby and relaxed object.
At the same time, 
it exhibits a number of past and current activities at core and larger scales, 
while at the cluster center, 
the very peculiar central galaxy NGC~1275 displays a complex network of emission-line nebulosity.
Based on the thread-like structure of filaments resolved by the Hubble Space Telescope,
\cite{Fabian2008} suggested the magnetic support of the structure
and ordered gas motions not highly turbulent over scales of $<10$~kpc around the galaxy.
The cluster X-ray emission peaks sharply toward the galaxy and its central active galactic nucleus (AGN), 
3C~84.
The AGN also develops jets and radio lobes 
where high-energy particles and X-ray-emitting-hot gas interact, 
as discovered by \cite{Boehringer1993}.
Associated with these current and possibly past AGN activities, 
fine gas structures such as bubbles, ripples,  and weak shock fronts 
were revealed by deep \chandra imaging \citep[][and references therein]{Fabian2011}.
These processes dissipate energies into  various types of gas motions.
Independently, based on the lack of resonant scattering of Fe-K line emissions, 
\citet{chura04} suggested gas motions in the core of a velocity of at least half the speed of sound.

West of NGC~1275, a chain of galaxies is distributed toward another radio galaxy IC~310, 
while 
other member galaxies and clusters virtually aligned with this chain
form the Pisces-Perseus supercluster.
On the other hand, the ICM X-ray emission has been traced up to at least 2~Mpc 
with an elongation in the east-west direction.
Based on the X-ray morphology of early observations, 
\cite{Hirayama1978} hypothesized that the gas and galaxies are rotating with a velocity of close to 3000 \kms .
\cite{Furusho2001} found asymmetrical ICM temperature distribution
and suggested a past merger in the direction parallel to the line of sight.
Using \xmm \citet{chura03} 
further studied the asymmetric structures around the core ($R<20'$) 
and  revealed  a hot ``horseshoe'' surrounding the cool core.
They  also assumed a minor merger along the east-west direction in the sky plane.
\cite{Simionescu2012} studied the gas distribution  over wider spatial scale ranges 
and suggested mergers and associated swirling gas motions.

The cluster center has been observed twice-yearly 
as a calibration target of the XIS
and we used all available data.
The integrated deep exposure of this unique object along with the high sensitivity of the XIS 
provides one of the best quality X-ray spectra from clusters.
We also used some offset region ($R<30'$) data and derived larger scale ICM emission properties.

Parts of the \suzaku Perseus data set were used for a number of measurements.
\citet{t09} reported the first detection of Cr and Mn X-ray lines from clusters and 
ICM elemental abundance measurements, 
 while \cite{Matsushita2013} and \cite{Werner13} reported abundance measurements out to the cluster outskirt.
Using offset XIS data along with the HXD data, 
\citet{nishino10} reported an upper  temperature limit in the outer region.
\citet{Simionescu2011,Simionescu2012} measured large-scale gas  thermal distribution over the viral radius.

In the next section,
we describe observations and data reduction, 
while in section~\ref{cal} we evaluate the accuracy of the XIS energy response, 
which is crucial to calibrate our measurement.
In sections~\ref{sect:ana-center} and \ref{ana-large}
we measure gas bulk velocities within the central pointing and over  a larger scale respectively, 
for the first time in this system, 
 which enables one of the most accurate measurements of gas bulk motion in relaxed clusters.
Finally these results are summarized and discussed in the last section.

Throughout this paper,
we assume the flowing cosmological parameters:
$H_0 = 70$ km s$^{-1}$Mpc$^{-1}$, 
$\Omega_\mathrm{m} = 0.3$, and $\Omega_\mathrm{\Lambda} = 0.7$.
At the cluster redshift of 0.0183, 
one arcmin corresponds to 22.2~kpc.
We use the 68\% ($1\sigma$) confidence level for errors, unless  otherwise stated.

\section{OBSERVATIONS AND DATA REDUCTION}
\label{sect:obs}

The Perseus cluster center is a calibration target
 and the central galaxy, NGC~1275, was always located at the CCD center.
We call these observations {\it central} pointings.
Here we use the data obtained from 2006 to 2013 taken in the normal window mode
and Table~\ref{obs:obs-center} shows an observation log.
The total exposure time is 576~ks.
The XIS~2 sensor was only available until November 2006.
These data were used in the instrument calibration reported by \citet{koyama07}, \citet{ozawa09}, and \citet{uchiyama09}.

\begin{figure}\begin{center}
\plotone{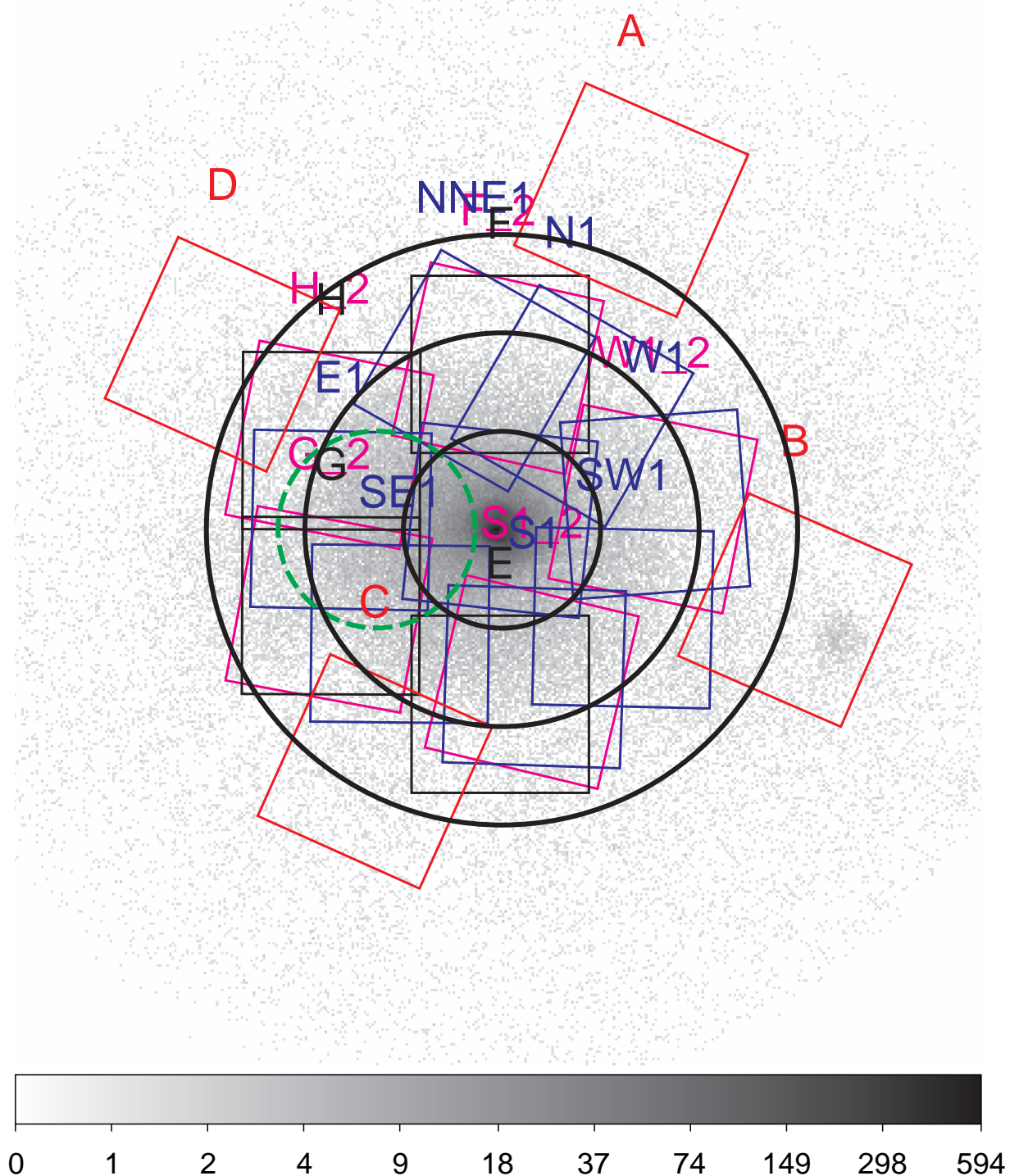}
\caption{\suzaku observation fields on top of the ROSAT image.
The XIS field of views
are shown in boxes.
Three circles indicate regions with radii of $10', 20'$ and $30'$.
The dashed circle indicates the east cool region found in \cite{Furusho2001}.
Observation-D is not used in the spectral analysis.
}
\label{obs:rosat2}
  \end{center}
\end{figure}

The off-center regions were also observed many times.
We call these {\it offset} pointings
 and use the data given in Table.~\ref{obs-offset}.
The total exposure time is 327~ks.
Fig.~\ref{obs:rosat2} shows  the field of views of the observations.

Detailed descriptions of the \suzaku observatory, 
 XIS instrument, 
and X-ray telescope  are found in 
\citet{mitsuda07}, \citet{koyama07}, and \citet{serlemitsos07}, respectively.

The XIS field of view is a square of $17'.8 \times 17'.8$ with $1024 \times 1024$ pixels.
Accordingly, 1 pixel $\simeq$ 1.04~arcsec.
The energy resolution is about 50~eV at 1~keV and 130~eV at 6~keV in FWHM.
Initially the XIS are operated normally in the spaced-row charge injection (SCI) off mode.
Since 2006 October, the SCI on has been used  in normal mode.

We started the analysis from archived cleaned event files, 
which were filtered with  standard selection criteria,
and used the latest calibration file as of 20012 November 6.

We examined light curves in the 0.3--2.0~keV band 
excluding the central bright region events ($R<6'$), 
 for stable-background periods.
None of the data showed any flaring event.

The XIS energy data have an original energy channel width of 3.65~eV.
We grouped these by common bin sizes of about one-third of the FWHM of the energy resolution.
Around the Fe-K line, 
the width is 8 channels or about 30~eV.
To establish the XIS energy response function, 
we used {\tt xisrmfgen} software 
alongside {\tt makepi} (version-20121009) to assign PI values.
The energy bin size is 1~eV ($\sim 0.15\times 10^{-3}$ of the Fe-K line energy), 
which means we can resolve the change of energy at less than the systematic limit of about $10^{-3}$.
For spectra fitting, we use XSPEC \citep[version-12.8; ][]{Arnaud1996} and 
maximum likelihood ($\chi^2$) statistics implemented in the same.
Note that the above binning gives about 15 counts per bin in the lowest-count spectrum and more than 20 counts per bin in other spectra analyzed below.

\begin{deluxetable}{llllcc}
\tablecaption{
\suzaku central pointings of the Perseus cluster. 
\label{obs:obs-center}
}
\tablewidth{0pt}
\tablehead{
\colhead{Name} & \colhead{Date} & \colhead{Sequence}    & 
\colhead{SCI}\tablenotemark{a} &
\colhead{Exp}\tablenotemark{b} & 
\colhead{Roll}\tablenotemark{c} \\
\colhead{} &  \colhead{}  &\colhead{}      & \colhead{} & \colhead{(ks)} & \colhead{(degree)} 
}
\startdata
CEN-0602  & 2006 Feb  & 800010010 & off & 43.7 & 260.2 \\
CEN-0608 & 2006 Aug & 101012010 & off/on & 92.0 & 66.0 \\
CEN-0702 & 2007 Feb & 101012020 & on & 40.0 & 258.7  \\
CEN-0708 & 2007 Aug & 102011010 & on &  35.1& 83.4 \\
CEN-0802 & 2008 Feb & 102012010 & on &  34.9 & 255.2  \\
CEN-0808 & 2008 Aug & 103004010 & on & 34.1 & 86.8  \\
CEN-0902 & 2009 Feb & 103004020 & on & 46.3 & 256.1  \\
CEN-0908 & 2009 Aug & 104018010 & on & 34.2 & 67.0  \\
CEN-1002 & 2010 Feb & 104019010  & on & 33.6 & 277.3  \\
CEN-1008 & 2010 Aug & 105009010 & on & 29.6 & 66.6  \\
CEN-1102 & 2011 Feb & 105009020 & on & 32.9 & 259.7  \\
CEN-1108 \tablenotemark{d}& 2011 Aug & 106005010 & on & 34.1 & 83.8  \\
CEN-1202 & 2012 Feb & 106005020 & on & 41.1& 262.0  \\
CEN-1208 & 2012 Aug & 107005010 & on & 33.2& 72.6   \\
CEN-1302 & 2013 Feb & 107005020 & on & 41.2 & 256.3  \\
\enddata
\tablenotetext{a}{Spaced-row charge injection mode on or off.}
\tablenotetext{b}{Exposure time.}
\tablenotetext{c}{Roll angle of the pointing defined as north to DETY axis. }
\tablenotetext{d}{No XIS-3 data is available.}
\end{deluxetable}

\begin{deluxetable}{lllllrrr}
\tablecaption{
\suzaku offset pointings of the Perseus cluster. 
\label{obs-offset}
}
\tablewidth{0pt}
\tablehead{
\colhead{Name} & 
\colhead{Date} & 
\colhead{Sequence}    & 
\colhead{SCI} & 
\colhead{Exp} &  
\colhead{R\tablenotemark{a}} & 
\colhead{PA\tablenotemark{b}} \\
\colhead{} & 
\colhead{} & 
\colhead{}    & 
\colhead{} & 
\colhead{(ks)} &  
\colhead{(arcmin)} & 
\colhead{(degree)} 
}
\startdata
 A & 2006 Sep & 801049010 & off& 25 & 36 & 345 \\
 B & 2006 Sep & 801049020 & off& 27 & 31 & 260 \\
 C & 2006 Sep & 801049030 & off& 31 & 28 & 140 \\
 D & 2006 Sep & 801049040 & off& 8  & 33 & 56 \\
 E & 2011 Feb & 805045010 & on & 27 & 18 & 180 \\
 F & 2011 Feb & 805046010 & on & 18 & 16 & 0 \\
 G & 2011 Feb & 805047010 & on & 17 & 14 & 110 \\
 H & 2011 Feb & 805048010 & on & 15 & 17 & 58 \\
E1       & 2009 Jul & 804056010 & on & 7  & 16   & 90 \\
N1       & 2009 Aug & 804063010 & on & 14 & 15   & 330 \\
S1       & 2010 Aug & 805096010 & on & 8  & 16   & 200 \\
W1       & 2010 Aug & 805103010 & on & 6  & 16   & 280 \\
SE1      & 2011 Aug & 806099010 & on & 11 & 15   & 140 \\
SW1      & 2011 Aug & 806106010 & on & 12 & 14   & 225 \\
NNE1     & 2011 Aug & 806113010 & on & 9  & 16   & 10 \\
F\_2      & 2013 Feb    & 807019010 & on & 14 & 16 & 0 \\
S1\_ 2     & 2013 Feb        & 807020010 & on & 23 & 16 & 200 \\
H\_2      & 2013 Feb        & 807021010 & on & 18 & 19 & 58\\
W1\_2     & 2013 Feb        & 807022010 & on & 23 & 16 & 280\\
G\_2      & 2013 Feb     & 807023010 & on & 14 & 19 & 110 \\
\enddata
\tablecomments{Observation D is not used in the spectral analysis, 
because of the low statistics.
}
\tablenotetext{a}{Distance between cluster and pointing centers.}
\tablenotetext{b}{Position angle of the pointing center with respect to the cluster center,
i.e. north, east, south, and west are 0, 90, 180, and 270 degrees, respectively. 
}
\end{deluxetable}

\section{ENERGY RESPONSE CALIBRATION}
\label{cal}
\subsection{Motivations}
\label{cal-mot}

The main goal of this paper is to measure Doppler shifts of cluster K-shell iron lines in  X-rays.
These astronomical shifts should be separated from 
instrumental effects on the energy scale (i.e. recorded pulse-height vs. energy).
The instrumental effects include
(i) CCD-to-CCD variation, 
(ii) intra-chip {\it spatial} variations  attributable to charge-transfer inefficiency (CTI)
and by CCD segment-to-segment qualities,
(iii) observation-to-observation {\it temporal} variation,
(iv) absolute energy determination, 
and unknown systematics.
These instrumental characteristics
have been calibrated and  extensively evaluated by the instrument team, 
which is an advantage of the \suzaku XIS 
over other CCD-type spectrometers.
These results have  also been demonstrated;
 not only in the cluster redshift analysis stated in section 1
but also in X-ray spectroscopy in the Galactic center and ridge \citep[e.g.][]{koyama07}
and in supernova remnants \citep[e.g.][]{Hayato2010}.
We summarize the latest calibration status in subsection~\ref{sect:cal-status}.
Using the latest calibration and the Perseus data, 
we further evaluate  accuracy in subsection~\ref{sect:cal-cal}.

Among the above instrumental variations,
factor (i) can be checked by comparing observed data from different CCDs in any observations.
The built-in calibration source lines are used to evaluate factors (iv) for each observation.

When analyzing the Perseus central data (section~\ref{sect:ana-center}),
factor (ii) is crucial
and (iii) can be internally checked against the data.
For the offset data analysis (section~\ref{ana-large}),
(iii) is important.
Moreover,  (ii) also affects the result because each pointing observation includes different distributions of events over the detector position.

\subsection{Reported Status}\label{sect:cal-status}
\citet{koyama07} estimated the systematic uncertainty of  absolute energy in the Fe-K band to be within $+0.1$ and $-0.05$\%, 
based on  Fe lines observed from the Galactic center alongside 
Mn K$_\alpha$ and K$_\beta$ lines (at 5895 and 6490~eV respectively) from the built-in calibration source ($^{55}$Fe).
Independently, 
OTA07 investigated the XIS data 
and evaluated the  energy-scale calibration in detail.
Using early calibration and early Perseus cluster data (sequence 800010010), 
they estimated the systematic error of the {\it spatial} gain non-uniformity (ii) 
to be $\pm 0.13$\% (68\% confidence level).
 A similar analysis was performed by \citet{tamura2011} using two Perseus cluster datasets with a new calibration
and the accuracy was confirmed.
Furthermore, 
\citet{ozawa09}  systematically examined the XIS data obtained from the start of  the operation in July 2005 until December 2006 (SCI off mode data).
They reported that  the {\it spatial} dependence of the energy scale (ii) 
 was well corrected for  charge-transfer inefficiency 
and  cited a time-averaged uncertainty of absolute energy of $\pm$ 0.1\%.
\citet{uchiyama09} reported  the energy scale calibration for SCI-on mode.
\citet{Sawada2012} reported an  improved CTI correction.
They archived energy accuracies of
$<0.7$\% and $<0.1$\% at 1~keV and 7~keV, respectively, over  the entire CCD position in the SCI-on mode data.
We based on these calibration methods in the latest calibration files.

The energy resolution  changed over time
and was measured at 140~eV (July 2005; SCI off),
200~eV (April 2007 ; SCI off),
and 160~eV (March 2008; SCI on) for the FI CCD at Mn K$\alpha$ in FWHM.
This gradual change in energy resolution was also calibrated;
the typical uncertainty of resolutions is 10--20~eV in FWHM.

\subsection{Calibration Source Data}
\label{sect:cal-cal}
To evaluate the absolute energy scale in different CCD segments (A and D in XIS-0,1,3)
we use the Perseus central pointing data (Table~\ref{obs:obs-center})
and extracted spectra of calibration sources which illuminate two corners of each CCD.
These spectra within the energy range 5.3--7.0~keV 
are fitted with two Gaussian lines (zero intrinsic width) for the Mn K${\alpha}$ and K${\beta}$ along with a bremsstrahlung continuum component.
We parametrize the energy scale using a common  ``redshift''
deviated from the expected value of 5895~eV (Mn K${\alpha}$) and 6490~eV (Mn K${\alpha}$).
Spectra from different observations, CCDs, and segments  were fitted separately.

 As shown in Table~\ref{cal:cal-var}, 
the average redshifts are all below $10^{-3}$, 
which indicates that the absolute energy scale is calibrated accurately
if averaged over observations.
The variations within each segment are comparable with statistical errors, 
which suggests that systematic errors in instrumental variations are below statistical ones.

Following OTA07 and  based on the obtained distribution of redshifts
we measure residual variation by 
calculating $\sigma_{\rm sys}$ such that $\chi^2 = \Sigma (z_i - <z>)^2/(\sigma_i^2 + \sigma_{\rm sys}^2)$
equals the degree of freedom.
Here $z_i, <z>$, and $\sigma_i$
are  the obtained redshift, its average, and statistical errors,
respectively.
We found $\sigma_{\rm sys}$ to be $0.8\times 10^{-3}$.
This variation includes CCD-to-CCD (i), 
segment-to-segment (ii), 
and temporal (iii) factors.
The calibration source is  attached to the opposite side of read-out nodes of each CCD;
 hence the above calibration source accuracy energy indicates the precision of the CTI calibration.

Averaging the four segments of the FI-CCDs (Table~\ref{cal:cal-var}), 
we obtain a redshift close to $-0.1\times 10^{-3}$, 
while BI redshift is close to $0.8 \times10^{-3}$.
This relatively significant BI spectral deviation, 
which is still within statistical errors, 
could introduce systematic error.
Because of better sensitivity of the FI at higher energy bands, 
FI is calibrated better than BI at these energy bands.
Moreover the FI spectra also provide better energy resolution. 
In view of these properties, 
we use only the FI data in the following analysis. 

To evaluate 
energy resolution calibration, 
we fitted the Mn lines by adding an artificial line width (\sigadd)
 and found that the best-fit \sigadd \, tended to be zero.
The 1 $\sigma$ upper limit of \sigadd \,  ranged from 5 to 80~eV depending on the line counts
with an average of 25~eV.
These confirm the reported 
 energy resolution calibration.
We assume this 25~eV $\sim 1250$ km s$^{-1}$ to be  a 1$\sigma$ systematic error 
in line width at around the Fe-K energy.

\begin{deluxetable}{lcccc}
\tablecaption{
Variation in the measured energy scale parametrized by redshifts over observations of the Perseus central pointing.
All values are in redshifts in units of $10^{-3}$.
\label{cal:cal-var}
}
\tablewidth{0pt}
\tablehead{
\colhead{CCD/seg\tablenotemark{a}} & 
\colhead{$<z>$\tablenotemark{b}} & 
\colhead{s.d\tablenotemark{c}} & 
\colhead{max-min\tablenotemark{d}} & 
\colhead{sta.err\tablenotemark{e}} 
}
\startdata
XIS0/A & 0.23 & 1.24 & 2.7 & 1.0\\
XIS0/D & 0.31 & 1.37 & 2.6 & 0.7 \\
XIS3/A & -0.11 & 0.63 & 1.6 & 1.1 \\
XIS3/D & -0.72 & 1.46 & 4.2 & 1.0 \\
\hline
XIS1/A & -0.65 & 1.28 & 3.5 & 0.9\\
XIS1/D & -0.95 & 1.43 & 2.8 & 1.4 \\
\enddata
\tablenotetext{a}{CCD and segment.}
\tablenotetext{b}{Average of redshifts.}
\tablenotetext{c}{Standard deviation. }
\tablenotetext{d}{difference between minimum and maximum values.}
\tablenotetext{e}{Average of statistical errors.}
\end{deluxetable}

\subsection{The Perseus Central Observations}
\label{cal-per}

\subsubsection{Motivations}

The spatial variation in the XIS energy scale [(ii) in subsection~\ref{cal-mot}] was calibrated 
using Perseus cluster data as one of the primal sources.
Here we use the same  source but all available data (Table~\ref{obs:obs-center})  alongside the latest calibration information and evaluate the accuracy.

We should note that the Perseus data themselves 
are used  to correct the CTI,
assuming  the cluster has uniform Fe-line energy over the central region ($17'.8\times 17'.8$).
Possible deviations from  this assumption could add systematic errors.
We assume these systematic errors to be insignificant  on the followings grounds.
Firstly, 
the CTI parameters are determined using  Perseus data 
observed in different roll-angles
, meaning the CTI correction is not solely dependent 
on cluster-intrinsic distribution of Fe-line energy.
Secondly, the CTI correction 
is not tuned for each observation,
but based on instrumental characteristics 
as a function of 
the number of CCD read-out transfer (ACT-Y coordinates), CCD segments, observation time, and X-ray energy.
Thirdly, 
built-in calibration source data
along with other astronomical sources such as the Galactic center and supernova remnants (at lower energies) 
were used to correct and evaluate the calibration.
We check the assumption more quantitatively in this subsection.

Here we group observations by their roll-angles.
One group (``winter'') data were obtained in February  each year 
with roll-angles of 255\arcdeg--277\arcdeg.
The other (``summer'') data were obtained in August with  roll-angles of 66\arcdeg--86\arcdeg.
Note that  pointing centers are close to each other within about 30 arcsec  in all data.
Within each group, 
different observations share roll angles close to each other, 
meaning a certain detector position allows a particular sky region to be observed.
Using these data  sets with two different extraction methods,
we  checked both {\it instrumental} and 
{\it cluster-}intrinsic variations in the energy scale.

In the first extraction (subsection~\ref{cel-per:det})
we  integrated spectra from different sky regions sorted by detector positions
and found no significant variation in the line centers measured among different detector positions
nor between the two-roll groups.
In the second extraction, given in section~\ref{sect:ana-center},
we integrated spectra from different detector positions sorted by sky regions.
We found no  significant variation among different sky regions
nor between the two-roll groups.
These  analyses are not completely independent 
and we cannot completely separate {\it instrumental} and {\it cluster} variations.
Nevertheless,  both indicate that both {\it instrumental} and {\it cluster} variations are small 
and within certain error level.

\subsubsection{Detector-sorted spectra}
\label{cel-per:det}
\begin{figure}[h]
\begin{center}
\epsscale{0.5}
\plotone{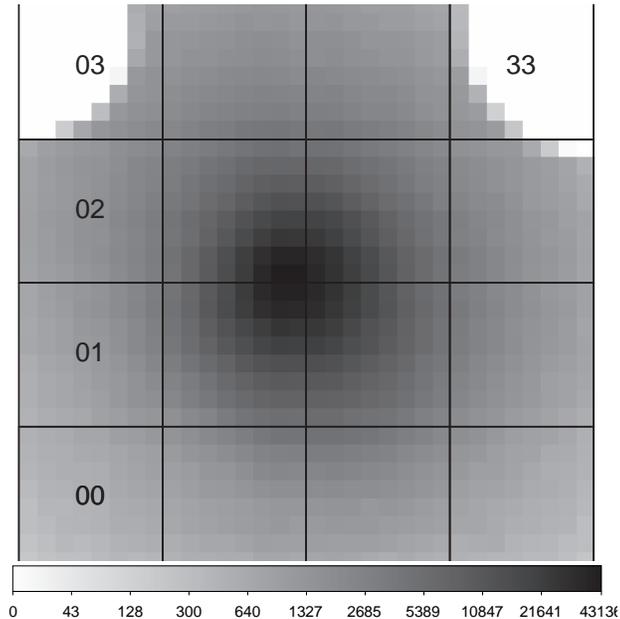}
\caption{Detector cells in (DETX, DETY) coordinates overlaid on 
the Perseus image of XIS-0 from one observation (101012010).
Parts of cell numbers are indicated.
The two top corners are regions illuminated by the calibration source and excluded for spectral extraction.
}
\label{cal:detector}
\end{center}
\end{figure}

We divided the XIS field of view into $4 \times 4$ cells of size $4'.5\times4'.5$
in detector coordinates (DETX and DETY) as illustrated in Fig.~\ref{cal:detector}.
We call these cells DET00, DET01 and so on up to DET33.

Based on each cell and observation, a spectrum is extracted, 
whereupon 
each spectrum in the 5.7--7.3~keV band is fitted with two Gaussian lines 
for He-like K${\alpha}$ ($\sim 6700$~eV) and H-like K${\alpha}$ ($\sim 6966$~eV)
and a bremsstrahlung continuum model.

As stated in subsection~\ref{ana-center-lw}, 
since we found no need for additional line width for Gaussian lines,
we fixed the width at zero henceforth.
We used the common redshift of the two lines as a fitting parameter.
Examples of the spectral fitting are given in Appendix (Fig.~\ref{cal-app:sp}).

\begin{figure}[h]
\begin{center}
\epsscale{0.5}
\plotone{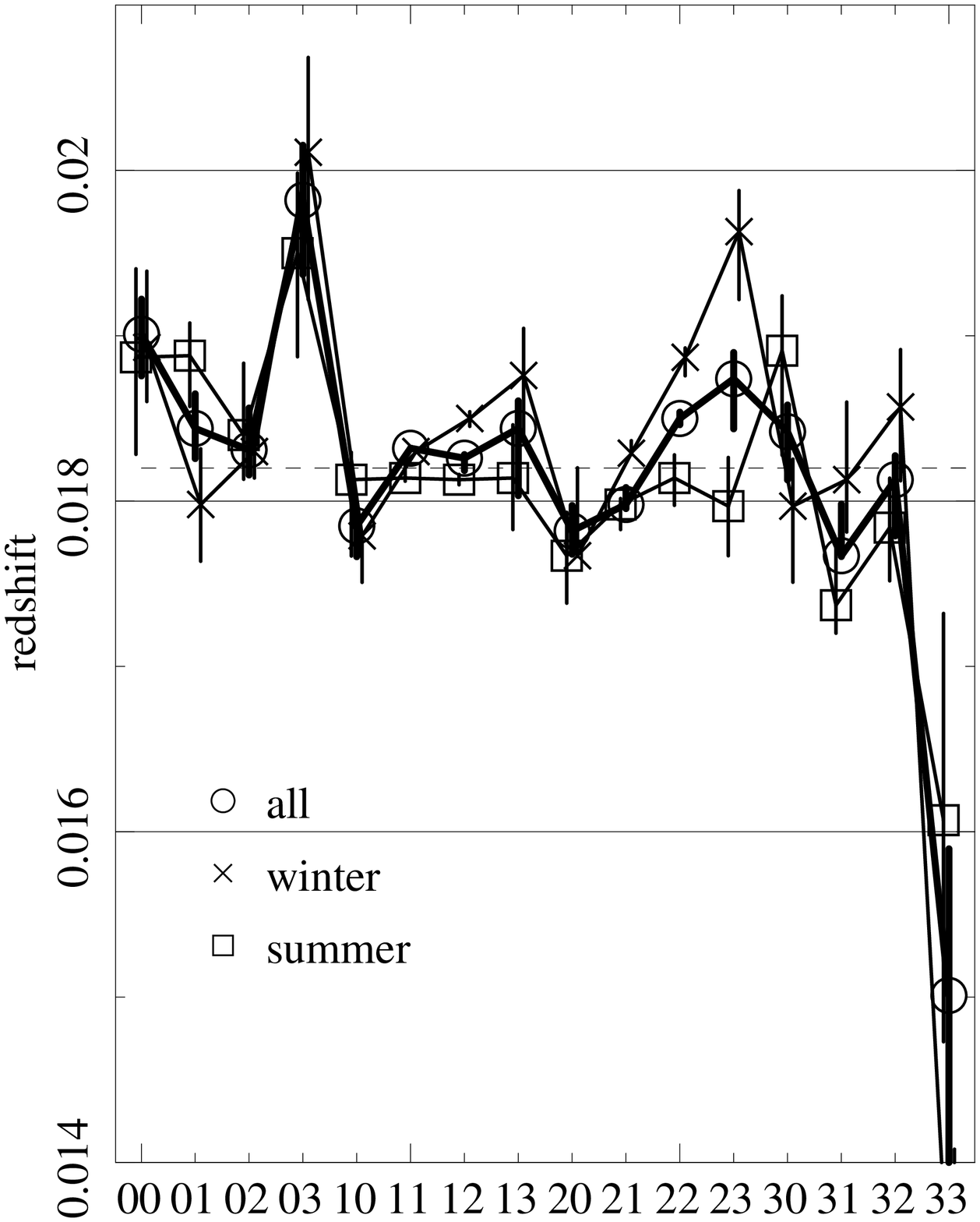}
\caption{Redshift (energy-scale) variations from spectra sorted by detector positions (X-axis).
Results from ``winter'' and ``summer'' roll-angles and all observations are given by different symbols.
}
\label{cal-per:reg-z}
\end{center}
\end{figure}

We found that 
the instrumental non-X-ray background is well-below the source count
and  subtracting the estimated background does not affect the obtained redshift, 
so we decided not to subtract the background from the central pointing data henceforth.
We confirmed that the two FI CCD (XIS-0 and -3) spectra
always give consistent line energies
 and this validates that CCD-to-CCD variation (i) is well calibrated,
at least within the FI CCDs.
We therefore combine these two CCD data  henceforth.
We fitted a set of spectra from individual cells (DET$ij$) from different observations within each roll-angle group using a common redshift.
We also fitted all observations using a common redshift.
The obtained redshift variations are given in Fig.~\ref{cal-per:reg-z} and Table~\ref{cal-per:det-var}.

There are two special cells, DET03 and DET33.
Because these include  calibration source areas, 
 they are much smaller in area (See Fig.~\ref{cal:detector}) and hence lower in counts than others.
These areas are not used here and in the Perseus spectral extraction given below.
In particular, the count of DET33 are lower than 10\% of those in others.
DET33  also shows the largest deviation from average, 
which could be  attributable to instrumental systematic effects but also  statistically.
DET03 shows larger redshifts  in both the two groups.
In this case, the deviation  is more likely caused by an instrumental effect.
These deviations in DET03 and DET33 are as small as $(1-3)\times 10^{-3}$.
In addition,  the sum of these two positions  occupies only a small percentage of the whole CCD area, 
which means any systematic errors have little effect  on our measurements.

We found that standard deviations, \sigsd ,
among  the detector positions for each roll-group and for all observations
are $(0.8-1.6)\times 10^{-3}$ in redshift (Table~\ref{cal-per:det-var}).
If we exclude the DET33 result, these are reduced to $\sim 0.6 \times 10^{-3}$.
These  variations are attributable to the combination of {\it instrumental} and {\it cluster} variations.

In the DET23 cell,
two-roll-group spectra  reveal the largest difference, $2.0\times 10^{-3}$, in redshifts, 
 but elsewhere, they give consistent redshifts.
Noting that different sky regions are observed by  both groups in the same detector position,
this indicates that {\it cluster-}intrinsic variation is  below these differences.

Further analysis of smaller spatial scales
is limited by statistics, 
but we assume that the instrumental energy scale does not change on such scales.

Based on these analysis,
we conclude that the CCD-to-CCD variation [(i) in subsection~\ref{cal-mot}]
and intra-chip {\it spatial} variation (ii)
 within spatial scales of a few arcminutes
is  below 0.1\% ($10^{-3}$ in redshift or 300 \kms \, around the Fe-K energy) over the whole CCD position.

\begin{deluxetable}{llccc}
\tablecaption{
Variations of measured redshifts from spectra sorted by detector positions.
All values are in redshifts in units of $10^{-3}$.
\label{cal-per:det-var}
}
\tablewidth{0pt}
\tablehead{
\colhead{roll-group} & 
\colhead{$<z>$} & 
\colhead{\sigsd \tablenotemark{a}} & 
\colhead{max-min \tablenotemark{b}} & 
\colhead{sta.err \tablenotemark{c}} 
}
\startdata
\multicolumn{5}{c}{All (16) positions }\\
winter & 18.2 & 1.6 & 5.9 & 0.39\\
summer  & 18.1 & 0.78 & 1.7 & 0.33\\
all & 18.2 & 1.0 & 3.4  & 0.23 \\
\hline
\multicolumn{5}{c}{Excluding DET33 (15 positions)}\\
winter & 18.5 & 0.66 & 0.91 & 0.28\\
summer  & 18.2 & 0.59 & 0.41 & 0.25\\
all & 18.4 & 0.56 & 0.60 & 0.18 \\
\enddata
\tablenotetext{a}{Standard deviation.}
\tablenotetext{b}{Difference between minimum and maximum values.}
\tablenotetext{c}{Average statistical error.}
\end{deluxetable}

\section{CENTRAL POINTING SPECTRA}
\label{sect:ana-center}
Here we use a set of central pointing data of the Perseus cluster 
and investigate the spatial distribution of spectral properties within the central region ($\sim 17'.6\times 17'.6$).

\subsection{Spectral Extraction}
\begin{figure}[h]
\begin{center}
\plotone{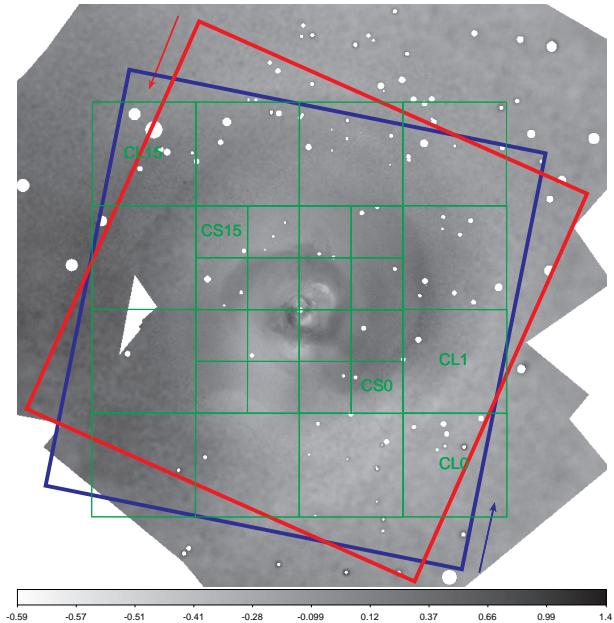}
\caption{Spectral extraction cells in (RA, DEC) coordinates (in green)
along with a \chandra image of the cluster.
The \chandra image shows the fractional difference to the average at each radius
taken from \cite[][Fig.2]{Fabian2011}.
Some of cell names are indicated.
North is up and west is right.
Two squares indicate XIS field of views by a single observation in the summer roll-angle group in red
and another in the winter roll-angle group in blue.
Arrows indicate the DETY direction from (DETX,DETY)=(0,0) corner for each observation.
}
\label{ana-center:regions}
\end{center}
\end{figure}

We divided the central sky region into cells of sizes $2'.2\times2'.2$ or $4'.4\times4'.4$,
as illustrated in Fig.~\ref{ana-center:regions}.
We call inner and smaller cells CS$i$ 
and larger cells CL$i$, 
with $i$ ranging from 0 to 15, 
starting from  the south-west corner toward  the north-east.
Note that  the sum of CL5,CL6,CL9,and CL10 overlap with CS0--CS15.
From each cell and each observation one spectrum is extracted. 

Before measuring redshifts,
we evaluate spectra by 
fitting  them in the 2.0--9.0~keV energy band with 
a single temperature model and the Galactic absorption (hydrogen column density of $1.5\times 10^{-21}$cm$^{-2}$)
 and determine the  temperature and metal abundance of each region.
We found a two-dimensional temperature variations of 4.5--6.5 keV.
Conversely, 
the metallicity increased radially from 0.3 to 0.45~solar toward the center.
These  results match previous measurements
and confirm that these spectra can be used to measure local properties from different sky regions.

\subsection{Redshift Measurements}
\label{ana-center-z}
\begin{figure}[h]
\centerline{\hbox{
\plotone{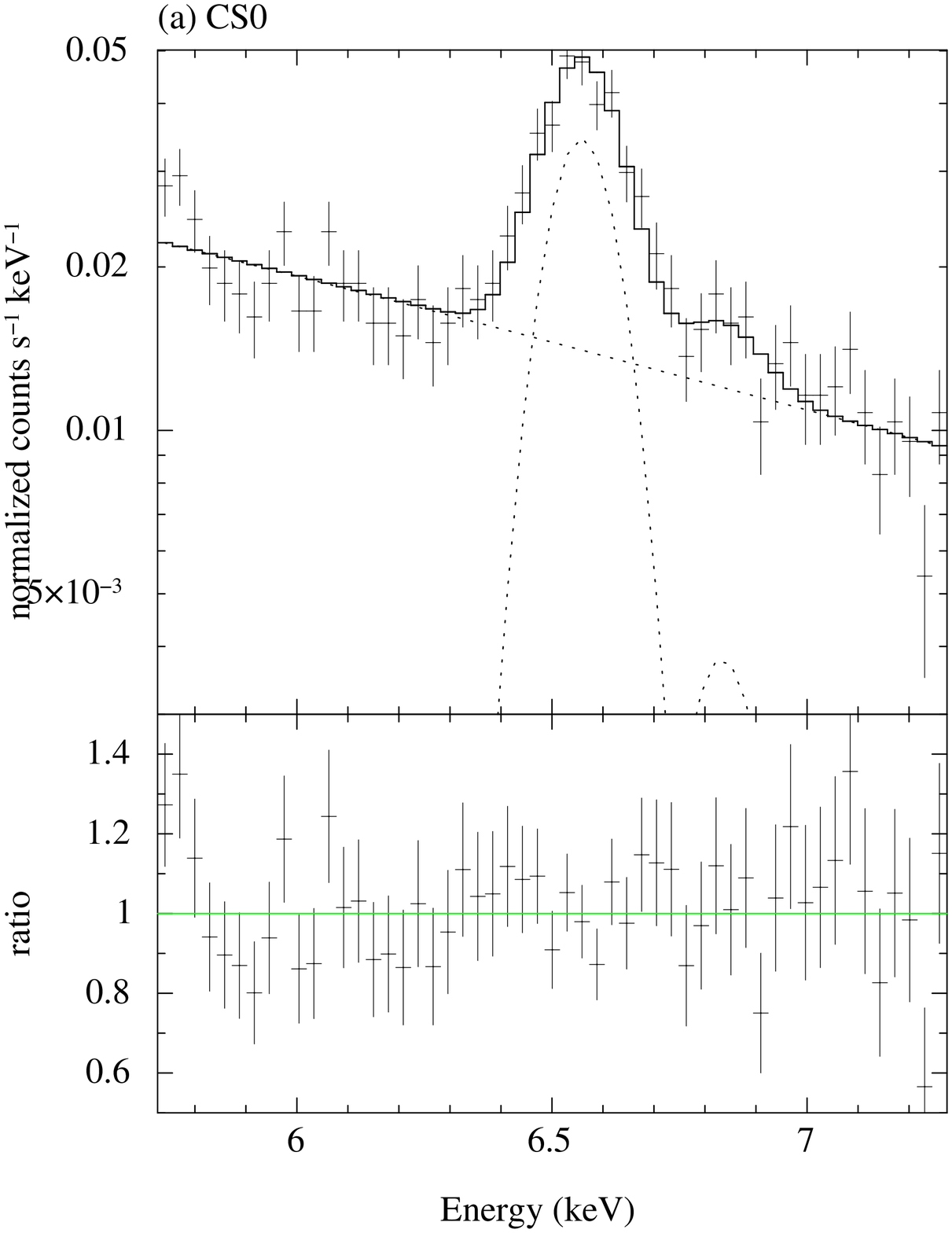}
\plotone{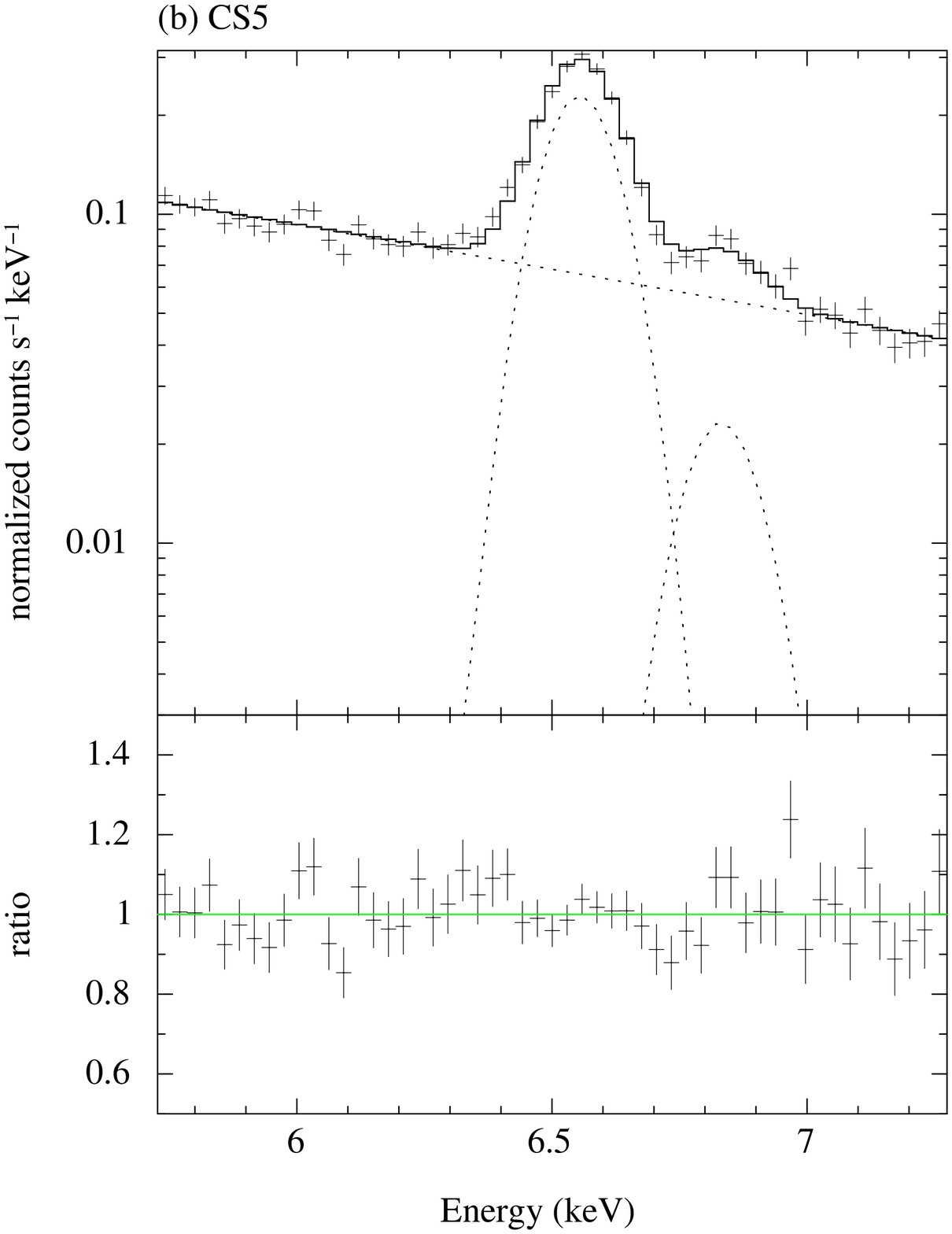}
}}
\centerline{\hbox{
\plotone{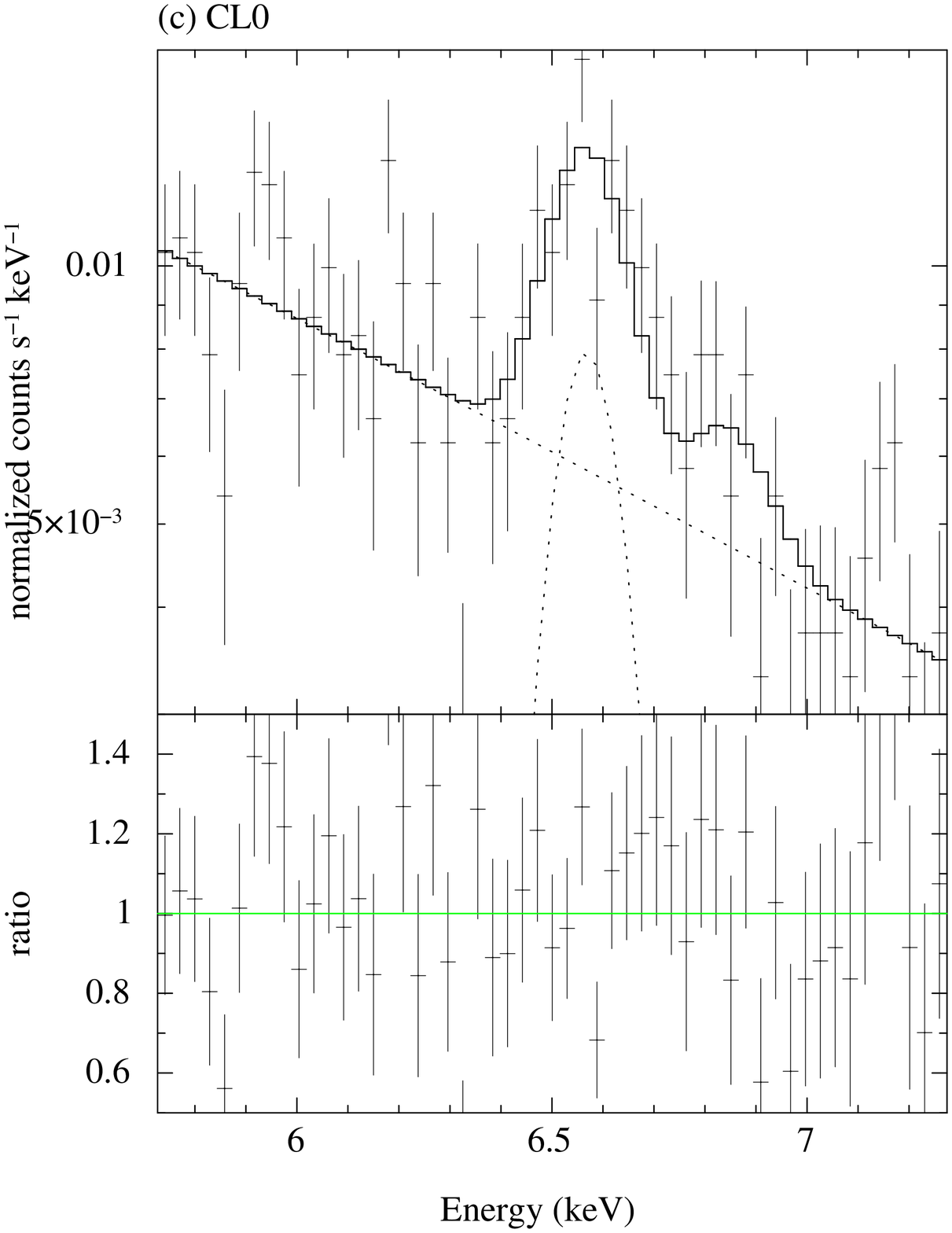}
\plotone{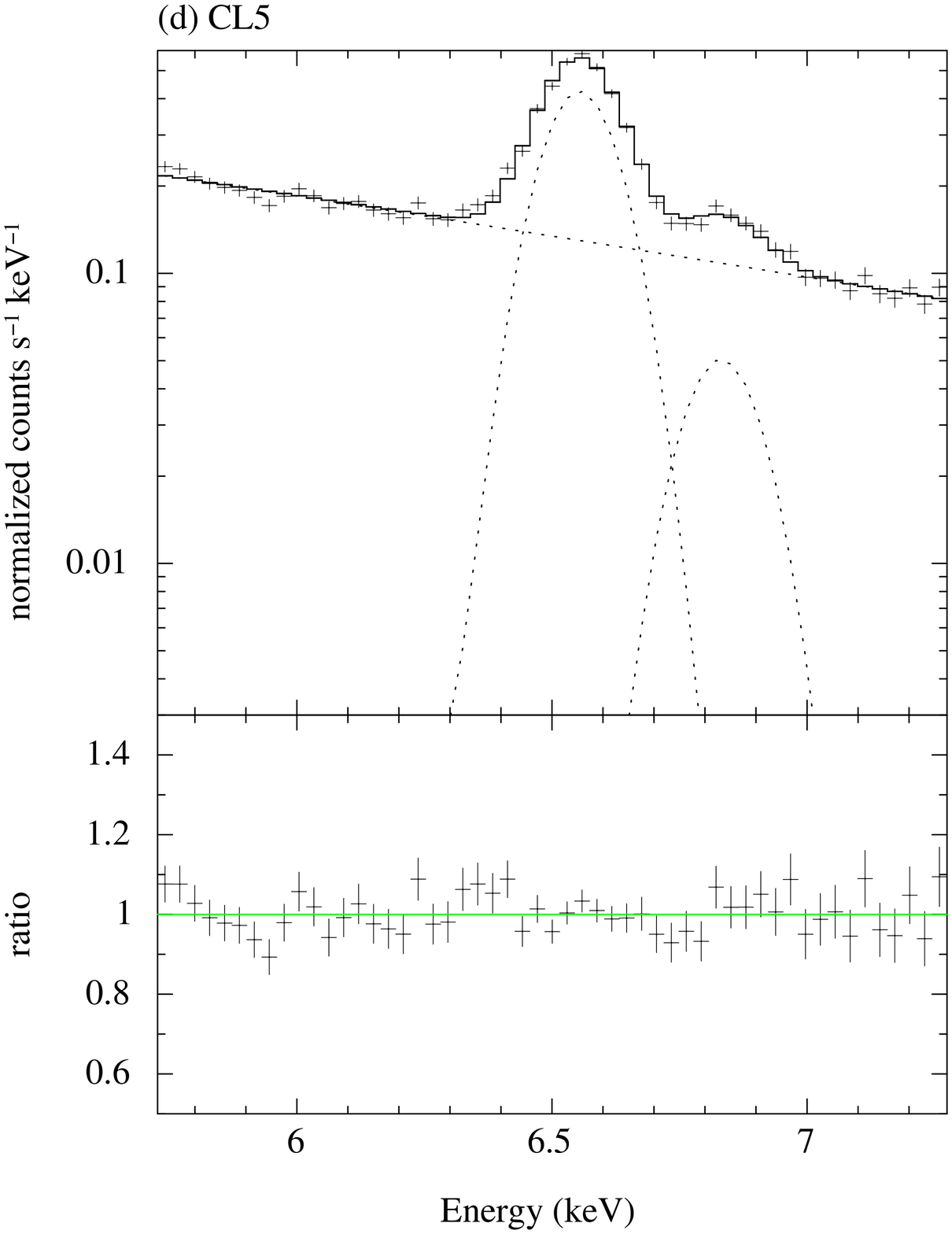}
}}
\caption{Example of the redshift determination.
The spectral fit from one sequence (104018010) of 
cells CS0, CS5, CL0, and CL5 is shown from the top left to the bottom right.
In the upper panels of each plot, 
the data, best-fit model, and model components are shown
by crosses, solid-histogram, and dotted lines, respectively.
In the lower panels fit residuals by the data to the model ratio are shown.
}
\label{ana-center-z:ex-sp}
\end{figure}

\begin{figure}[h]
\plotone{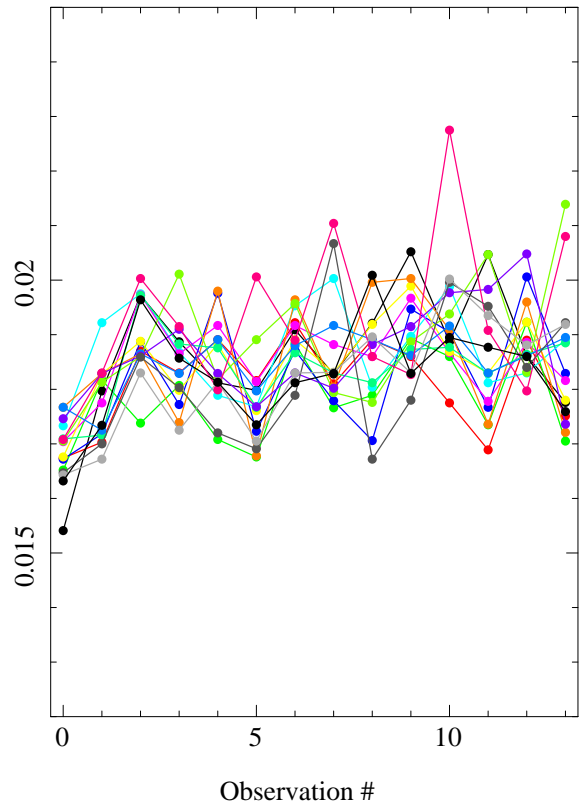}
\caption{
Obtained redshifts as a function of observation time (sequences) for different sky regions (CS0 to CS15).
Statistical errors are about $\pm 0.5\times 10^{-3}$ and $\pm 0.8\times 10^{-3}$ on average for inner and outer cells, 
respectively.
}
\label{ana-center-z:seq_z}
\end{figure}

We measure redshifts based on the Fe-K line center
and the same model (two Gaussians and a bremsstrahlung continuum) as in section~\ref{cal-per}.
Fig.~\ref{ana-center-z:ex-sp} shows  fitting examples.
The redshift variation as a function of observation and hence observation date
is shown in Fig.~\ref{ana-center-z:seq_z}.
Spectra from the first observation (sequence 800010010) show systematically smaller redshift.
Other than this, however,
there is no systematic change over the observation period, 
confirming that the {\it temporal} variation in the energy scale is effectively corrected.

\begin{figure}[h]
\centerline{\hbox{
\plotone{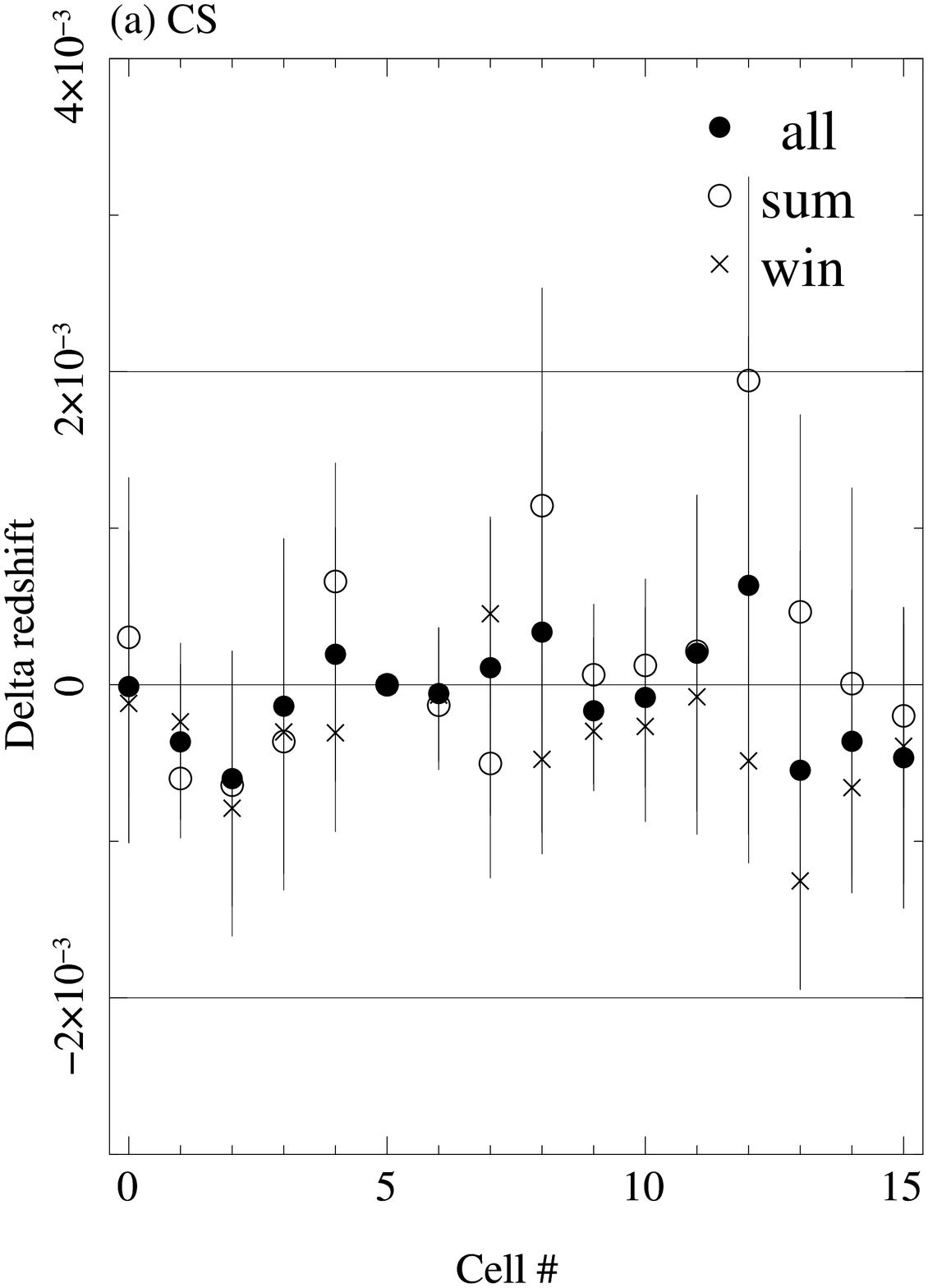}
\plotone{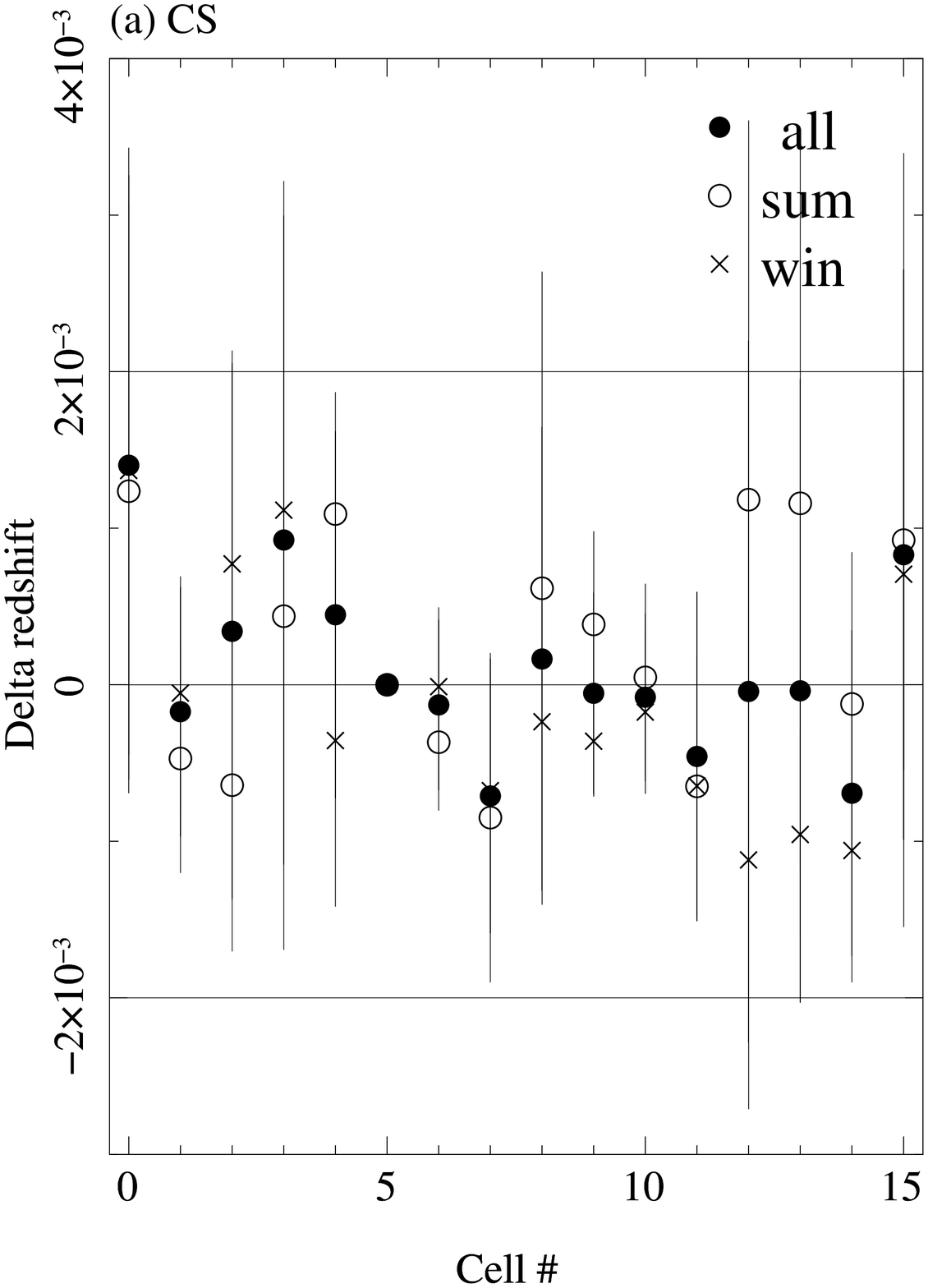}
}}
\caption{
Obtained redshifts as a function of sky regions.
Left and right panels show results of inner cells (CS0 to CS15)
and outer cells (CL0 to CL15), respectively.
Error bars are shown only for all observations.
}
\label{ana-center-z:reg_z}
\end{figure}

To compensate possible observation-dependence of the energy scale,
we measure  the redshift relative to that of a bright cell (CS5 or CL5) for each observation and each sky cell, $\Delta z$. 
Fig.\ref{ana-center-z:reg_z} shows $\Delta z$ averaged over observations.
Here we measure  the error (\sigsdall) for each cell 
 based on standard deviation among all observations.
These observations differ in  terms of the detector orientation (i.e. roll-angle) and observation-date.
Therefore the scope of error includes not only statistical errors
 but also at least partially systematic errors
due to instrumental energy-scale variations.
Along with average values from all observations,
those from summer and winter roll-angle observations are given separately.
Differences between the  two-roll-angle observations enable 
systematic errors from instrumental effects to be estimated.
In all cases,  however, the differences are below $2\times 10^{-3}$, 
confirming that the {\it instrumental} variation is insignificant as described in section~\ref{cal-per}.

The largest $|\Delta z|$ is $1.5\times 10^{-3}$ at CL0, 
where \sigsdall \, is about $1.9\times 10^{-3}$.
All other regions
have $|\Delta z|$  below $1.0\times 10^{-3}$.
Averages and standard deviations of the best-fit $\Delta z$ along with 
averages of \sigsdall \, are given in Table~\ref{ana-center:rms}.
These results show  uniform $\Delta z$ distribution 
and hence no gas motion  exceeding these variations and errors.
 The errors for each sky region range from $\pm 0.5\times 10^{-3}$ to $\pm 2\times 10^{-3}$,
corresponding to radial velocities of 150 and 600~\kms \, respectively.

\begin{figure}[h]
\centerline{\hbox{
\plotone{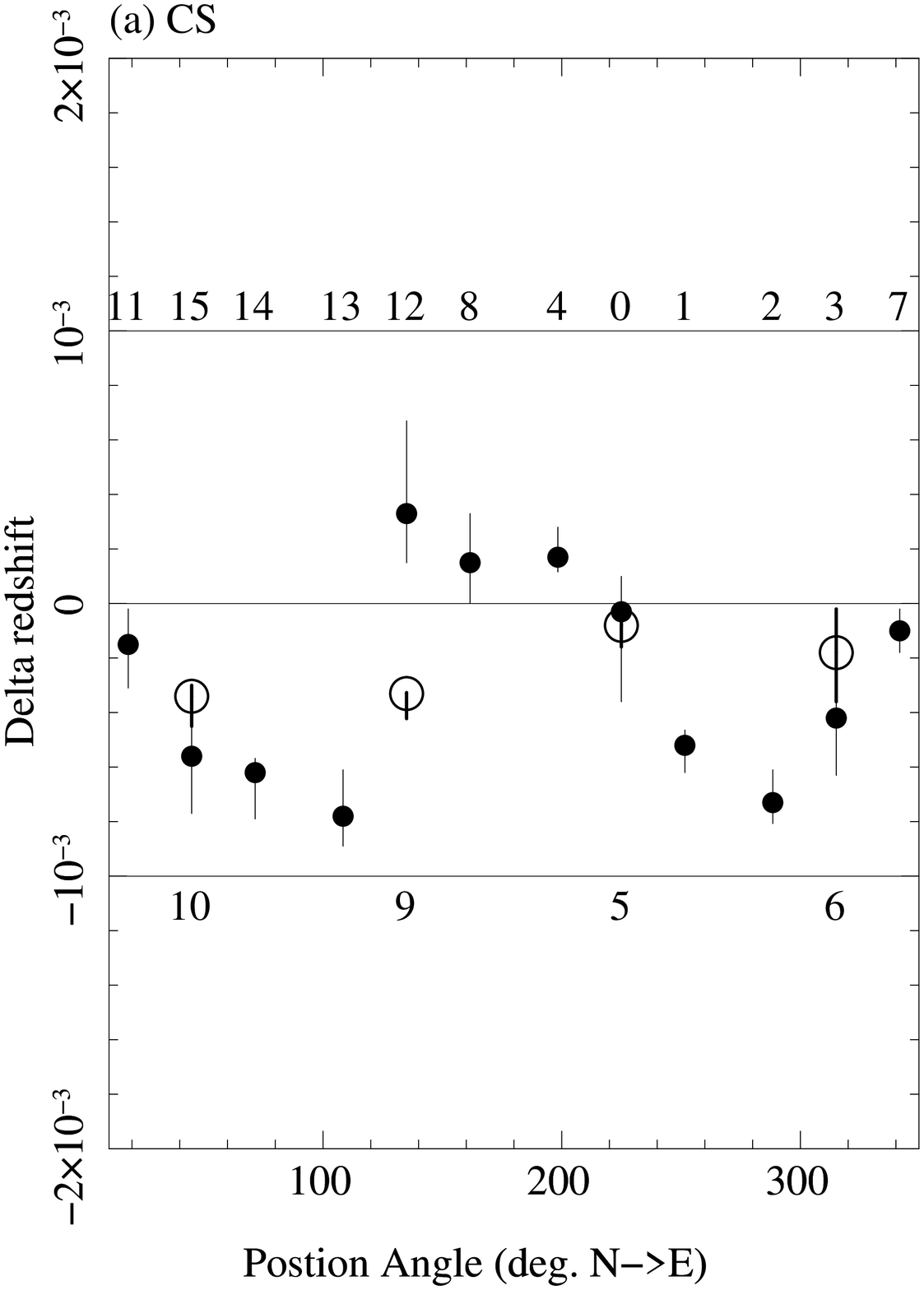}
\plotone{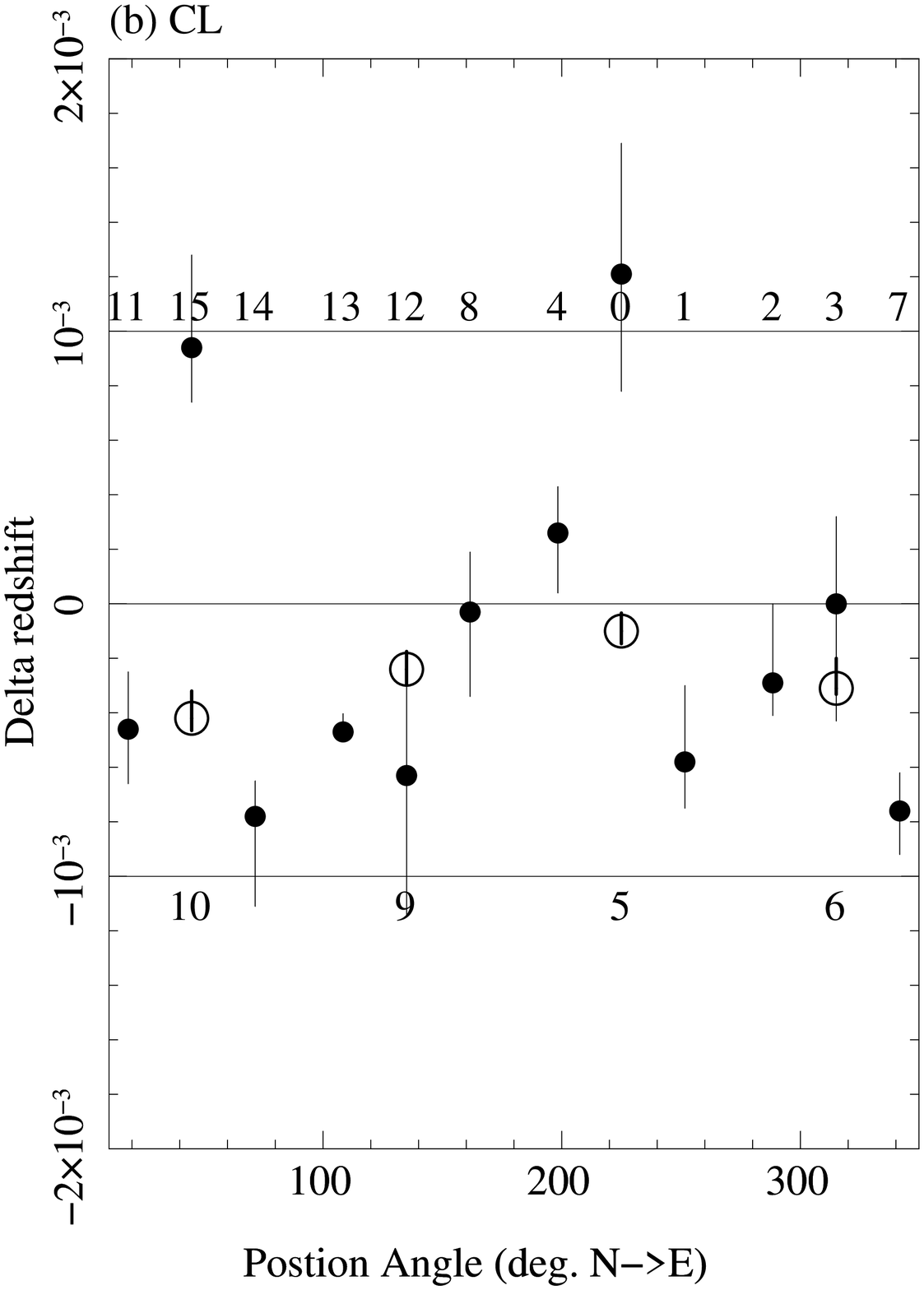}
}}
\caption{
Obtained redshifts as a function of position angle (north to east).
Panels (a) and (b) show
results of smaller cells (CS0--CS15)
and from larger cells (CL0--CL15), respectively.
Filled- and open-circles
are from the outer twelve cells and inner four cells, respectively, for each region.
Numbers in panels indicate cell numbers.
}
\label{ana-center-z:pa_z}
\end{figure}

Within the inner cells, 
CS1--CS3 show lower $\Delta z$ in the best-fit value.
To further search for possible spatial variation  in redshifts,
we use  data differently from the above case.
Here we fit all spectra for each cell simultaneously with a common redshift, 
assuming no instrumental and systematic variation in the energy scale.
We remove one observation (sequence 800010010) to avoid systematic error  as mentioned above.
We sorted the result as a function of  the position angle of the cell
with respect to the cluster center, as shown in Fig.~\ref{ana-center-z:pa_z}.
The obtained redshift are basically consistent with the above result in Fig.~\ref{ana-center-z:reg_z}., 
 although the given error declines because these include only statistical errors.
From the inner cells (left panel Fig.~\ref{ana-center-z:pa_z}) 
we found lower redshifts in  easterly and westerly directions
and higher redshifts  toward the south.
From the larger cells (left panel Fig.~\ref{ana-center-z:pa_z}) 
we found some variations but more  random
and comparable to the possible systematic error 
(section~\ref{cal-per}; $\sim 10^{-3}$ in redshift or 300 \kms ).
We  cannot find all variations if we use the data in summer or winter observations separately 
(See Fig.~\ref{ana-center-z:reg_z}).
Based on these careful analyses, 
we suggest a hint of lower redshift regions at $2-4'$ (45--90~kpc; CS1--CS3) west of the cluster center.
Other variations are  insignificant and possibly  attributable to instrumental effects.

In a search for redshift variation  on a finer scale,
we divided the central $3'\times 3'$ region
into 9 cells of  sized $1'\times 1'$.
We found a uniform redshift distribution within about $\pm 1\times 10^{-3}$ in redshift.

\begin{deluxetable}{lccc}
\tablecaption{
Statistics of redshift variations among 16 different sky regions.
All values are in unit of redshift in $10^{-3}$.
\label{ana-center:rms}
}
\tablewidth{0pt}
\tablehead{
\colhead{regions} & 
\colhead{$<\Delta z>$} & 
\colhead{\sigsd \tablenotemark{a} } & 
\colhead{$<$err$>$ \tablenotemark{b} } 
}
\startdata
CS0--CS15 & -0.1 & 0.35 & 0.92\\
CL0--CL15 & 0.12 & 0.52 & 1.3 \\
\enddata
\tablenotetext{a}{Standard deviation.}
\tablenotetext{b}{Average of errors.}

\end{deluxetable}

\subsection{Line Broadening and Turbulent Velocity}
\label{ana-center-lw}
The observed spectra are used to constrain the turbulent Doppler broadening of the emission line.
Following OTA07 and similar to our analysis  stated in subsection~\ref{sect:cal-cal}, 
we model the Fe-line spectra by including an additional line width (\sigadd; a Gaussian width).

Here we note that the  cluster-intrinsic lines are not a single Gaussian line,
but integrate many lines.
For example  the emission around 6.7~keV includes not only a dominant He-like resonance line at 6700~eV
but also intercombination lines at 6667-6682~eV 
and a forbidden line at 6636~eV
along with other weak lines between the resonance and forbidden lines.
Effective widths for this He-like triplet are about 30~eV for  the temperature of 2--4~keV gas (OTA07).
In addition, gas bulk motions  unresolved spatially within the spectral extraction region and projected in the line of sight contribute to \sigadd.

\begin{figure}[h]
\begin{center}
\centerline{\hbox{
\plotone{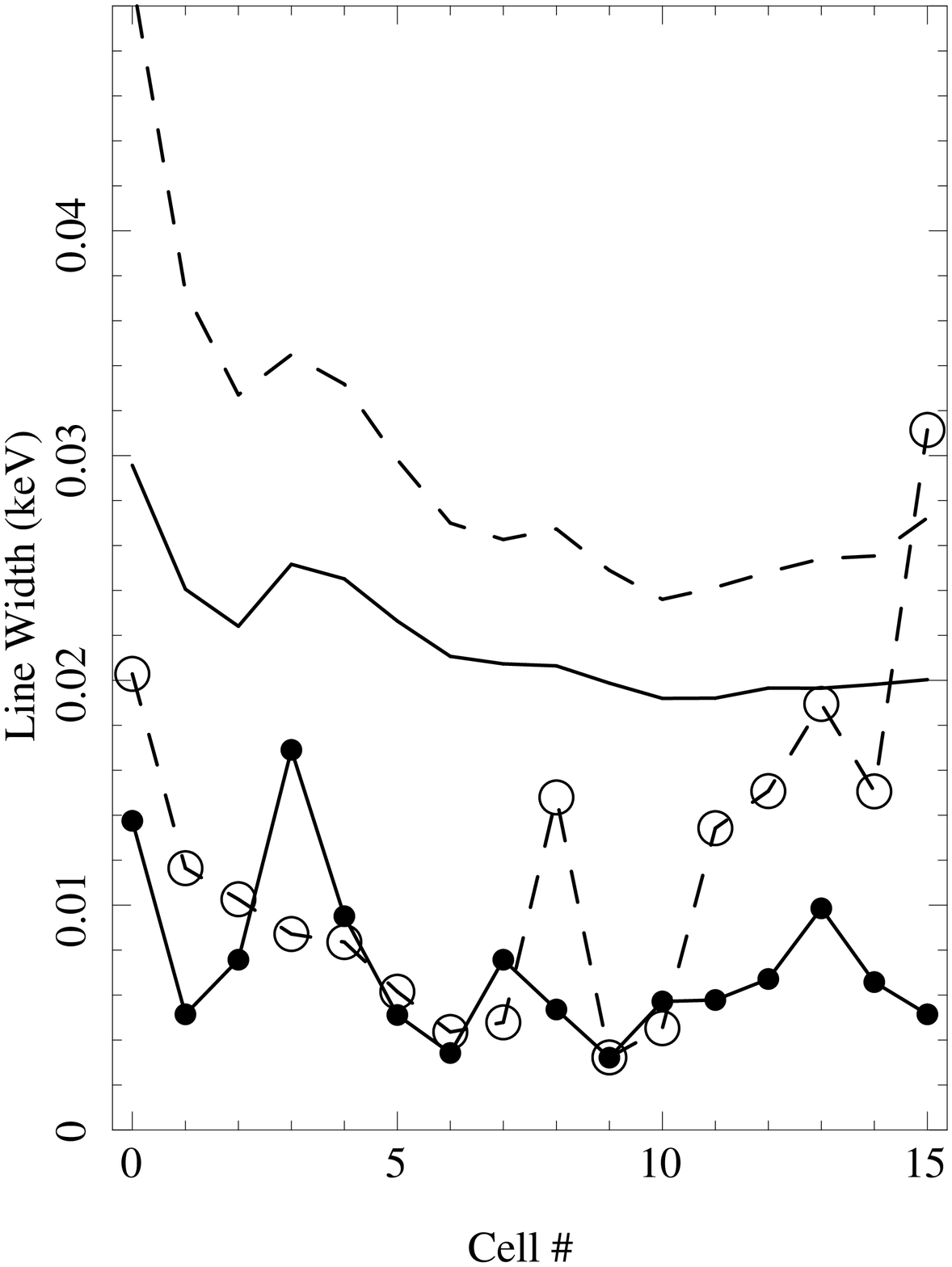}
}}
\caption{
Additional line widths, \sigadd (circle marks), 
and their upper limits (line) from different sky regions, 
averaged over different observations.
Results from smaller cells and larger cells
are shown by solid and dashed-lines, respectively.
Note that at CL15 
\sigadd \, exceeds its upper limit,
because both are average values from different observations.
}
\label{ana-center-z:lw}
\end{center}
\end{figure}

For each observation and region, 
we obtained \sigadd \, and its upper limit.
We found that in most cases \sigadd \, was consistent with zero and that even using a non-zero \sigadd \, did not change the best-fit redshift.
These are averaged over different observations and presented in Fig.~\ref{ana-center-z:lw}.
The best-fit \sigadd \, are all  below 20~eV, except for that from CL15 where  the statistics are significantly lower than others.
Given the systematic error and intrinsic cluster line width, 
these \sigadd \, are all consistent with no additional broadening due to turbulent motion.
Statistical upper limits of \sigadd \, are 20--50~eV, 
 which equates to the upper limits for turbulent motion of 900--2200 \kms.
We may add systematic uncertainty of 1250 \kms (subsection \ref{sect:cal-cal}).

\section{LARGE-SCALE SPECTRA}
\label{ana-large}

Here we use a set of offset pointing data with offset angles of $14'-36'$ from the cluster center
(Table~\ref{obs-offset} and Fig.~\ref{obs:rosat2}) 
and study the spatial redshift distribution  on a large scale.

We extract a spectrum from the entire CCD field of view for each offset observation.
Here we use the same Gaussian model as in subsection~\ref{ana-center-z}
to fit the FI 5.7--7.0~keV band spectra and measure the redshift distribution.

\begin{figure}
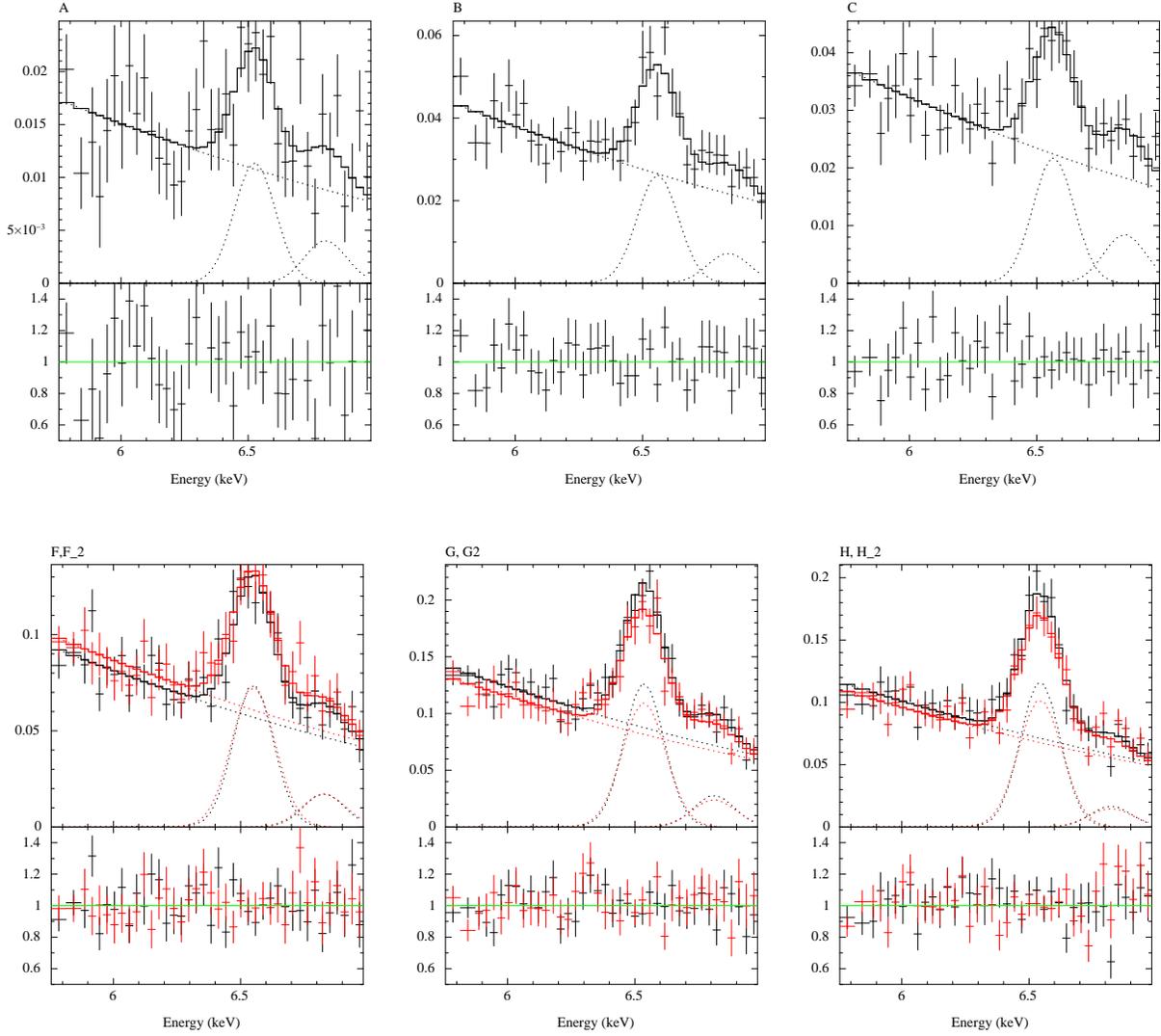

\begin{center}

\includegraphics[height=0.35\textheight]{fig10a.ps}
\includegraphics[height=0.35\textheight]{fig10b.ps}
\includegraphics[height=0.35\textheight]{fig10c.ps}
\includegraphics[height=0.35\textheight]{fig10d.ps}
\includegraphics[height=0.35\textheight]{fig10e.ps}
\includegraphics[height=0.35\textheight]{fig10f.ps}
\caption{
Example of spectral fitting from offset pointing data,
which are used to measure redshifts.
From top left to bottom right,
the spectra from regions
A, B, C, F, G, and H are shown.
For each plot, 
in the upper panels
the data, best-fit model, and model components 
in counting units of s$^{-1}$ keV$^{-1}$ are shown.
In the lower panels fit residuals in terms of the data to model ratio are shown.
}
\label{ana-large-z:ex}
\end{center}\end{figure}

There are some sets of two observations ( X and X\_2)
sharing a pointing direction but with different roll-angles.
We found that these data provide consistent results within the set, 
 hence in these pointing sets, spectra are fitted simultaneously with a common model.
Examples fittings are shown in Fig.~\ref{ana-large-z:ex}.

\begin{figure}[h]
\begin{center}

\includegraphics[scale=0.4]{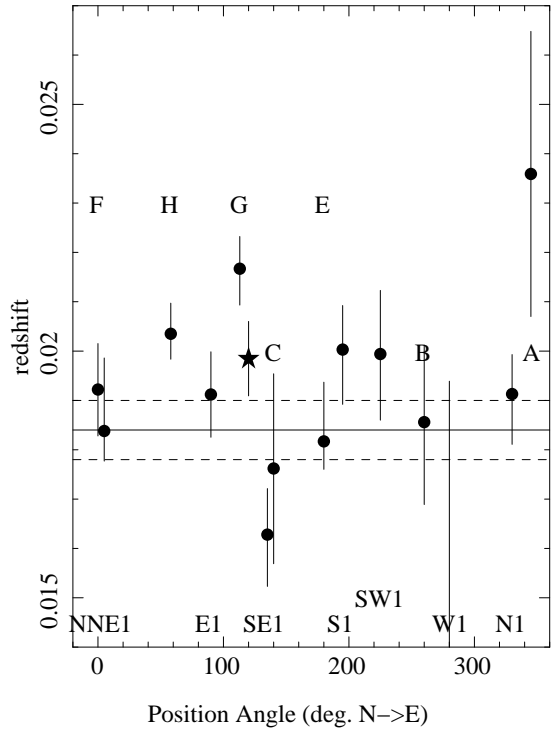}
\caption{
Redshift distribution from offset observations as a function of position angles of the pointing.
The star mark shows the redshift obtained by a simultaneous fit of G and SE regions.
The horizontal solid and dashed lines
indicate the average and standard variation from the central pointing (subsection~\ref{ana-center-z}).
}
\label{ana-large-z:z1}
\end{center}
\end{figure}

The obtained redshifts are shown in Fig.~\ref{ana-large-z:z1}.
We notice that  G and SE regions show larger and smaller redshifts, respectively, 
compared with the average.
These two regions, however, are  in proximity 
each other and overlap in  the sky regions by about half the area of the entire field of view.
Therefore, 
these deviations are likely statistical or systematic errors, 
and  do not originate entirely from the cluster gas redshift variation.
When we combine these two regions, a redshift of $0.020\pm 0.001$ 
consistent with other regions, is obtained, 
as shown in Fig.~\ref{ana-large-z:z1} (a star mark).
Region A also shows a relatively large deviation  alongside significant statistical errors.
Including these deviated regions, 
 the error-weighted average, standard deviation, and average statistical error
are found to be 
$19.2 \times 10^{-3}$,
$1.8 \times 10^{-3}$, and
$1.2 \times 10^{-3}$, 
respectively.

Based on the above results,
we conclude that gas redshifts are uniform within $\pm (1-2) \times 10^{-3}$ 
over the cluster core region ($R<20-30'$) in a spatial scale of about $15'$ (300~kpc).
In other words, 
we found no systematic velocity structure  exceeding $\pm (300-600)$~km s$^{-1}$.

The average and standard deviation 
of redshifts in the central region given in subsection~\ref{ana-center-z})
are $18.4 \times 10^{-3}$
and $0.6 \times 10^{-3}$, respectively.
The redshift averaged over the offset regions is consistent with that of the central region within these variations and errors.

To constrain line broadening and turbulent velocity,
we performed a similar spectral fitting as in subsection~\ref{ana-center-lw}.
However, 
lower statistics mean
the limits obtained are typically weaker than those in the central regions above.
We found the obtained \sigadd\, to be consistent with zero turbulent velocity
and statistical upper limits of \sigadd\, of 20--70~eV, 
corresponding to velocities of 900--3000 km s$^{-1}$.

\clearpage
\section{SUMMARY AND DISCUSSION}

\subsection{Spatial Mixing}
\label{dis-limit}
The photon mixing caused by the \suzaku telescopes
and projection effects
meant the measured velocity variation
was diluted compared with the intrinsic three dimensional variation.

The point spread function of the telescopes  include half-power diameters of $1'.8-2'.3$
and 90\% fraction encircled-energy radii of $2'.5-3'.0$,
meaning 
variations below these scales would be smoothed.
It is also difficult to separate the gas velocity of a faint emission component. 
This effect was examined by a ray-tracing simulation in OTA07
for the \suzaku measurements in the Centaurus cluster.
They found that when $2'\times 2'$ cells are used at the core, 
about 50-60\% of the  events collected by each cell
 originate from the surrounding eight cells.
Because the Perseus cluster has a central surface brightness distribution
similar to that of Centaurus, 
our measurement in the central region with $2'.2$ width cells (CS0 to CS15)
 is similarly affected. 
For the outer cells with a $4'.4$ width (CL0 to CL15), 
mixing effects should be much smaller ($<10$ \%).

In the large-scale measurements (section~\ref{ana-large}),
photon mixing due to stray X-rays from the bright center could be an additional contamination.
We estimate, however, that photons scattered from the central region to offset regions with a typical offset angle of  $15-20'$
 represent a minute percentage of all photons within each offset pointing
and  we could hence ignore this mixing.

\subsection{Summary of Results and Comparison with Previous Studies}
\label{dis-com}

\cite{dupke01-per} claimed a radial velocity difference in the Perseus cluster using the ASCA GIS. 
They compared spectra collected at the center and offsets separated by more than $40'$.
These spatial scales  exceed our study within radii of $30'$,
 which means these two measurements cannot be compared.

We summarize  the limits we determined for gas motion
along with related parameters in Table~\ref{dis-com:summary}.
For comparison,  results of OTA07 are also presented.
Both results of the \suzaku XIS are limited by similar levels of systematic errors
from the energy scale calibration.
The levels of statistical errors between the two are also comparable.

As described in section 1, 
gas velocity measurements are important to calibrate 
the hydrostatic equilibrium (H.E.) mass estimation.
OTA07 converted  velocity variation obtained into rotational motion
and approximated the uncertainty of the H.E. mass
by  gas kinetic to thermal energy ratio.
The ratio is proportional to $f M^2 = f (\frac{v}{c_s})^2$, 
where $f$, $v$ and $M$ are 
the fraction, velocity, and Mach number of the moving gas
and $c_s$ is the gas sound velocity (given in Table~\ref{dis-com:summary}).
The uncertainty of the rotation angle introduces another error.

Our limits on gas bulk velocities and hence on $M$
are comparable to those in OTA07.
The wider physical size covered by our observations (row R in Table~\ref{dis-com:summary})
allowed us to constrain the gas motion and hence the total mass within a volume significantly exceeding that in OTA07.
This is an important step to measure total masses in cluster scales.
Provided the Fe-K line emission of the Perseus central region  is exclusively 
brighter than other nearby clusters, 
this system could  represent a unique measurement of the gas motion at the cluster scale via current CCD-type instruments.
The decreasing X-ray surface brightness  hinders efforts to constrain the motion beyond our achieved scale.
It will  also remain challenging to reach out beyond cluster core regions
even with near-future high energy resolution instruments
such as SXS \citep{mitsuda10} onboard \astroh \citep{takahashi10}
without a large collecting area. 

\begin{deluxetable}{llccc}
\tablecaption{
Summary of gas velocity limits.
\label{dis-com:summary}
}
\tablewidth{0pt}
\tablehead{
\colhead{} & \colhead{} & \colhead{Perseus} & \colhead{Perseus} & \colhead{Centaurus} \\
\colhead{} & \colhead{}  & \colhead{(center)} & \colhead{(large)} & \colhead{} \\
\colhead{} & \colhead{(unit)}  & \colhead{section~\ref{sect:ana-center}} & 
\colhead{section~\ref{ana-large}} & 
\colhead{OTA07}
}
\startdata
redshift & & 0.0183 & -- & 0.0104 \\
$1'$ & (kpc) & 22.2 & -- & 15.3 \\
$kT$ \tablenotemark{a} & (keV) & 4.5  & 6 & 3 \\
$c_s$ \tablenotemark{b} & (\kms) & 1100 & 1300 & 900 \\
region size \tablenotemark{c}& & $2'.2/4.4'$ & $18'$ & $2'.1$ \\
& (kpc)       & 50/100 & 400 & 30\\
R  \tablenotemark{d} & (kpc) & 200 & 600 & 150 \\
$\Delta v_{\rm bulk}$ \tablenotemark{e}& (\kms) & 300--400 & 550 & 700 \\
 &                      & $\pm 300$ (sys.) & $\pm$ 300 (sys.) & $\pm$500 (sys.) \\
\vturb \tablenotemark{f} & (\kms) & 900--2200 & 900--3000 & 900\\
 &    (\kms) & $\pm$ 1250 (sys.) & $\pm$ 1250 (sys.) &  $\pm$ 1250 (sys.)\\
\enddata
\tablenotetext{a}{Typical ICM temperature within the measured region.  
The Perseus temperatures are taken from spectroscopic result in \citet{t09}, while that of the Centaurus from OTA07.
}
\tablenotetext{b}{Sound speed at the ICM temperature.}
\tablenotetext{c}{Individual region sizes.}
\tablenotetext{d}{Averaged separation among regions.}
\tablenotetext{e}{Upper limit of the velocity variation.
Systematic error is indicated by 'sys'.
Limits are taken from the standard deviation from samples.}
\tablenotetext{f}{Upper limit of the turbulent velocity.}
\end{deluxetable}

\subsection{Dynamics in the Core}
\label{dis-core}

In and around NGC~1275, 
complex interactions occur among 
various gaseous components, high-energy particles from the AGN 
and  radiation from stars and AGN, and probably magnetic fields.
Nearby galaxies as well as the central galaxy  may have moved around  within the growing gravitational potential.
In this active environment, the ICM should have various types of motions.
Observationally, the lack of resonant scattering of  the Fe-K line at the Perseus core indicates gas motions of $M>0.5$ 
\citep{chura04}.
 It is crucial to measure the energies in these gas motions in our study.

We found a hint of gas bulk motions within the cluster core (Fig.~\ref{ana-center-z:pa_z}; $R<4'.4$), 
while the key detection was lower velocities in regions $2-4'$ west of center.
West of center, clear X-ray enhancement was found as seen in Fig.~\ref{ana-center:regions}.
\cite{chura03} interpreted this structure as a result of 
a past minor merger in  an east-west direction. 
The obtained gas motion is consistent with this minor merger model
and  additionally suggests a velocity component in the line of sight.
Because we observe a mix of photons from the sub component and  main cluster
due to the sky projection and the telescope point spread function,
the real gas velocity of the sub  exceeds the observed figure ($v_{\rm obs}=150-300$ \kms).
If we assume $f$ to be  the emission fraction from the sub,
$v_{\rm obs}$ can be approximated with sub and main velocities ($v_{\rm sub}$ and $v_{\rm main}$) 
as $v_{\rm obs} = f v_{\rm sub} + (1-f) v_{\rm main}$.
In this case, the relative velocity  with respect to $v_{\rm main}$
can be estimated as $\Delta v_{\rm sub} \sim f^{-1} \Delta v_{\rm obs}$.
From the relative deviation map of the surface brightness (Fig.~\ref{ana-center:regions}),
we estimate $f$ to be 0.2--0.4.
In this case, 
$\Delta v_{\rm sub}$ could be 500--1000 \kms.
Interestingly, the second optically-bright member galaxy, NGC~1272, 
sitting on a rim of the west sub structure ($\sim 5'$ WWS of NGC~1275), 
also has a lower radial velocity than that of NGC~1275 by $\sim 1100$ \kms.
Furthermore, the chain of galaxies further west 
has an average radial velocity lower than the cluster center by about 1500 \kms.
We suggest  the association of the obtained gas velocity structure with these galaxies. 
We discuss these gas and galaxy relations  on a larger scale in the next subsection.

Using the obtained measurements, 
we estimate kinetic energy ($E_{\rm kin}$) by  gas bulk motion.
As stated in section~\ref{dis-com}, $E_{\rm kin}$ relative 
to the thermal energy ($E_{\rm th}$) in the gas 
can be presented by 
$f M_{\rm bulk}^2 = f (\frac{\Delta v_{\rm bulk}}{c_s})^2$.
Using the relation between 
$\Delta v_{\rm sub} \rightarrow \Delta v_{\rm bulk}$ 
and $\Delta v_{\rm obs}$ given above,
$E_{\rm kin}/E_{\rm th}$ becomes $f^{-1} (\frac{v_{\rm obs}}{c_s})^2$.
Assuming $c_s$ of 1100 \kms \, and $f$ of $0.2-0.4$, 
the observed $\Delta v_{\rm obs}$ (150--300 \kms) gives
$E_{\rm kin}/E_{\rm th} = 0.1-0.3$.
\cite{Fabian2011} estimated the energy associated with non-radial structure ($E_{\rm d}$)
based on  variation in the gas pressure map.
This energy should be closely related to $E_{\rm kin}$ estimated here.
They estimated  $E_{\rm d}/ E_{\rm th}$ to be within a few percent at radii of $\sim 110$~kpc 
corresponding to the west sub component region.
Because both estimations are order of magnitude, 
we  cannot compare these quantitatively.
These estimations are among the first attempts to measure all the energy components in the cluster gas
and understand the energy distribution of clusters.

The west system is the  largest  substructure within the core.
Our measurement  also indicates that it has the largest bulk velocity,
at least in a radial direction, 
 which means this structure could have the largest kinetic energy due to bulk motion.
We found no other systematic motions around the core with scales $>20$~kpc and 
bulk velocities  exceeding half the sound velocity ($\sim 550$~km s$^{-1}$).
This is consistent with  the observation that any other features revealed around the core
are  below 10~kpc ($30''$) in size.
These structures  are also likely to have velocities 
not exceeding the virial velocity of the central galaxy, or about $200$~km s$^{-1}$.
 The lack of heated gas or strong shock \citep{Fabian2003} also pointed  to no sonic or super sonic motion.
We conclude  that within much of the cluster core volume, 
the gas remains in hydrostatic equilibrium.

In Table~\ref{dis-cd:z} we present radial velocities of NGC~1275 and the cluster from optical data  and this X-ray results.
The optical radial velocity of NGC~1275
differs from that of a mean of cluster member by only about 200 km s$^{-1}$,
demonstrating that the galaxy  remains at the bottom of the cluster potential.
The absolute radial velocity of the gas at the core
is consistent with that of the galaxy in optical within the errors from X-ray data.
This indicates that large parts of stars and the gas in and around the galaxy
stay together at the bottom of the cluster potential center.
The cluster gas  on a large scale also has a mean velocity consistent with that in optical, 
confirming that these two baryonic components beyond galaxy scale  remain together at the same potential center.
These comparisons in absolute redshifts are among the first attempts  involving clusters.

Gas dynamics in cluster cool cores has been studied by numerical simulations.
For example, \cite{Asca06} simulated off-axis minor mergers and gas sloshing and reproduce cold fronts.
They presented evolutions of gas velocity structures.
Compared with large velocity variations in their representative runs,
our measured variations  (Figs.~\ref{cal-per:reg-z} and ~\ref{ana-center-z:pa_z})
are more uniform spatially and close to their results for late stages 
\citep[e.g. 6/4.8 Gyr in Fig.5/10 of ][]{Asca06}.
Note that our Doppler measurements are not sensitive to motions in the plane of the sky and diluted by projection.
In fact in the Perseus cluster gas spiral flows in the plane of the sky are suggested from X-ray images 
\citep[e.g.][]{Fabian2011, Simionescu2012}.
Focusing on AGN jet-driven bubbles, 
\cite{Heinz10} simulated high-resolution X-ray spectroscopy of nearby clusters including the Perseus cluster.
They found the velocity structure of the bubbles in this core with energy shifts of about 350 \kms, 
which are close to our limit.
However, these structures are too small in size ($<10$ arcsec) and too faint in X-ray brightness contrast to resolve by the current data.

\begin{deluxetable}{lccl}
\tablecaption{
Radial velocities of galaxies and X-ray gas. 
\label{dis-cd:z}
}
\tablewidth{0pt}
\tablehead{
\colhead{target}        &  
\colhead{velocity} & 
\colhead{error} & 
\colhead{Reference}\\
\colhead{}        &  
\colhead{(km~s$^{-1}$)} & 
\colhead{(km~s$^{-1}$)} & 
\colhead{}
}
\startdata
NGC~1275      & 5250 & 20 & 1 \\ 
member galaxy mean   & 5470 & 100 & 2  \\ 
X-ray gas at center & 5520 & 300 & section~\ref{sect:ana-center} and \ref{ana-large}\\ 
X-ray gas at $R<15'$ & 5760 & 600 & section~\ref{ana-large}\\ 
\enddata
\tablerefs{(1) NED database; 
(2)\cite{kent1983}.
}
\end{deluxetable}

\subsection{Large-Scale Dynamics}

We found an absence of gas bulk motion  on a large scale.
Our upper limit  for the radial gas velocity is about 600~km s$^{-1}$, 
 which implies 
either that gas motion in the cluster is predominantly in the plane of the sky or subsonic at most.

In subsection~\ref{dis-core} we discussed  the potential of the gas motion around the western core being associated with galaxy motions.
The associated chain of galaxies  is further distributed toward west.
 To study the gas and galaxies relation in the west and larger areas,
we compare our derived gas dynamics with that in galaxies in the cluster.

\begin{figure}
\begin{center}
\includegraphics[scale=0.8]{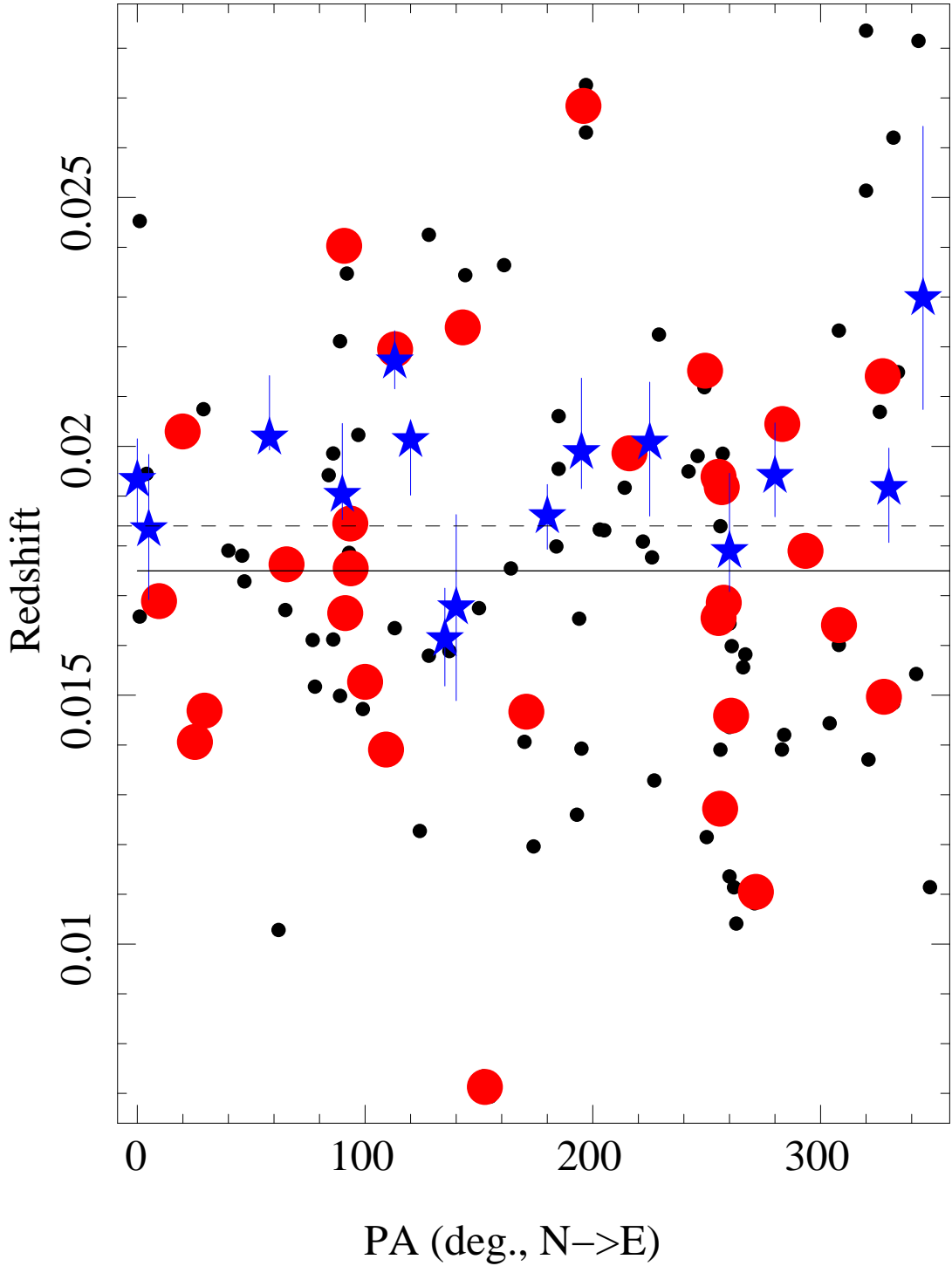}
\caption{
Radial velocity distributions of galaxies and gas as a function of the position angle.
The small (black) and large (red) circles indicate galaxies from the CfA redshift catalog (94 galaxies)
and bright galaxies (K-band magnitude $<10.7$ ) from the 2MASS redshift catalog.
There are some overlaps between the two samples.
The velocity of NGC~1275 is shown by a solid line.
Star marks indicate the gas velocity taken from Fig.~\ref{ana-large-z:z1}.
The central velocity of the gas components is shown by a dashed line.
}
\label{dis-large:pa-vh}
\end{center}
\end{figure}

We use the following two catalogs
as the deepest collections of galaxies in the Perseus cluster.
The CfA redshift catalog \citep{Huchra1995}
includes 94 (120) galaxies with radial velocities
within $30'$ ($60'$)  of the cluster center, 
while the 2MASS redshift survey \citep{Huchra2012}
includes 66 (103) galaxies in the same regions but with complete photometric data.
Fig.~\ref{dis-large:pa-vh} shows  the radial velocities of galaxies 
within a radius of $30'$  of the center 
as a function of azimuthal angles  relative to the center.
Most galaxies  are distributed around the cluster dynamical center at redshift of $\sim 0.018$, 
corresponding to that of NGC~1275.
This is consistent with early studies of galaxy distributions \citep[e.g][]{kent1983}.
Some galaxies have higher or lower velocities ($\Delta v > 2000$ km s$^{-1}$), 
which may be contamination  dynamically unrelated to the main system.

We include gas velocities from Fig.~\ref{ana-large-z:z1} into Fig.\ref{dis-large:pa-vh}, 
 which is the first such comparison.
In galaxy velocities, there are some local structures, 
 the clearest of which
is a group at an angle of around 260\arcdeg \, with lower velocities 
($-2000$ km s$^{-1} < \Delta v < 1000 $ km s$^{-1}$).
 This is part of the chain of galaxies mentioned in subsection~\ref{dis-core},
which should be before or after a collision into the main cluster  centered on NGC~1275.
Based on the X-ray distribution around the core, 
\cite{chura03} suggested that the substructure is after a collision about 0.25 Gyr ago.
 Based on comparison between the galaxy and gas velocities around this angle at the cluster west,
we notice that the gas is closer to the cluster center than galaxies at this azimuth.
Note that these galaxies in total exceed the cD galaxy in terms of luminosity and hence mass.
If these two baryonic components are associated,
what  explains this difference?
We note a difference in effective radii between the two wavelength bands.
The X-ray emission is weighted toward the inner region, 
 due not only to the positions of spectral extractions (Fig.~\ref{obs:rosat2})
but also the density-squared X-ray emission intensity.
Considering these effects, 
the effective radius of the X-ray data can be approximated  around $10'$.
On the other hand, galaxies in the west are uniformly distributed over the $30'$ radius region.
These galaxies are point sources in X-ray without  any bright extended excess emission \citep{Fabian2011}.
In addition,
this difference suggests a segregation between the two components during a violent collision.
Such detachments in the sky position have been observed in some merging systems such as the Bullet \citep{Clowe06}.
Our measurement is a new approach to reveal such actions not only in spatial  terms but also in the radial velocity space.
Note that in the major-merger system Abell~2256 
\citet{tamura2011} revealed that gas and galaxies move  in pairs within radial velocity and spatial spaces.

Based on ASCA observations 
\cite{Furusho2001} observed 
an extended cool region $10-20'$ east of the cluster center, 
 alongside a ring-like region surrounding the east cool region and the core.
They suggested a past collision of a poor cluster in a direction nearly parallel to the line of sight.
Our observation covers a substantial fraction of the east cool region as shown in Fig.~\ref{obs:rosat2}.
To limit the current velocity of the east sub system 
we  must assume the X-ray emission fraction
of the moving system within the extracted spectra.
We estimate this  at 0.3--0.5 according to the relative deviation map of surface brightness in \cite{chura03}.
Combining this with our measurement in Fig.~\ref{ana-large-z:z1} ($\Delta v <600$ \kms),
we constrain the radial velocity of the sub system $<(1200-2000)$ \kms.

In future we would  cover more galaxies by  conducting deeper surveys
and deep X-ray spectra covering larger volumes in a number of systems.
Gravitational lensing would also measure dark matter distributions.
Direct comparisons between these three components
in dynamical and spatial distributions
should reveal a number of local events and measure the dynamical cluster age,  
 thus giving a systematic picture of structure formation of gas and galaxies but invisibly controlled by dark matter.

Considering that the virial radius ($r_{\rm vir}$) of the cluster is 2.2 Mpc 
(or $\simeq$ 100'), we measured the gas motion typically around $R \simeq 0.1-0.3 r_{\rm vir}$. 
The obtained upper limits for the bulk motions are typically 
$(\Delta v_{\rm bulk}/c_s) \lesssim 0.5$ taking account of the systematic errors. 
The turbulent motions are more loosely constrained ($v_{\rm turb} \gtrsim c_s$).
Cosmological cluster simulation results suggest that the kinematic energy of the gas 
is expected to be left only in the level of $\sim 10$ \% of the
thermal energy in the corresponding radius range 
\citep[e.g.][]{Lau09, Vazza09}. 
Furthermore, this value tends to be even lower for a sub sample of relaxed clusters. 
Therefore, our results of the absence of both the bulk and turbulent motions except for the
some hints at west of the core are consistent with these simulations.

\subsection{Turbulent Motion}

In addition to the gas bulk motion,
we searched for the turbulent motion using the line width.
We obtained the limits generally larger than the sound speed 
as given in Table~\ref{dis-com:summary}.
OTA07 also reported a limit of the turbulence
using the XIS data of the Centaurus integrated over a large region of $18' \times 18'$ (Table~\ref{dis-com:summary}).
These two limits are similar to each other,
because that both limits depend largely on the energy resolution of the XIS and its calibration.
Our result is the first attempt to constrain the turbulent motion based on the spatially-resolved line X-ray spectra.

The gas turbulent (random) motion could introduce an error on the H.E. mass estimate.
The additional mass or kinetic energy due to the turbulence can be approximated by the factor given in subsection~\ref{dis-core}, 
$M^2$ with the turbulent velocity, \vturb \, replaced as $v$.
We obtained the limit on \vturb \, a few times larger than that on the bulk velocity (Table~\ref{dis-com:summary}).
Therefore the uncertainty from the turbulent motion dominates that from the ordered one.

The \xmm RGS was used to constrain the gas turbulent motions in hot gas in clusters and galaxies
\citep[e.g.][]{Xu2002}.
\citet{Sanders2013} analyzed RGS spectra of dozens of sources 
and found evidence for $>400$ \kms \, velocity line broadening and limits down to 300 \kms \, for some systems.
The RGS spectra can be used only for centrally-peaked X-ray line emission below 1.5~keV (dominated by Fe-L lines) 
and results are coupled with the brightness distribution of the line emission which is not straightforward to model.

As described in subsection \ref{ana-center-lw}, 
the limit on \vturb \, is mostly limited by the instrumental energy resolution.
Therefore to improve the limit significantly high energy resolution instruments such 
as X-ray calorimeters are necessary.
For example, 
direct and robust study will be obtained by SXS onboard \astroh,
providing 5--7~eV (FWHM) energy resolution, 
corresponding to 300 \kms \, at the Fe-K line.
A 100 ksec SXS observation of the Perseus center
provides more than 10,000 counts in the He-like triplet emission 
and measurements of the turbulent along with those of bulk velocity structure.

\vspace{1cm}
We thank the referee for useful comments and suggestions.
\chandra image of the Perseus cluster was provided kindly by J.Sanders.
We thank all the {\it Suzaku} team member for their supports.
We acknowledge the support by a Grant-in-Aid for Scientific Research from the MEXT, 
No.24540243 (TT), No.25400231 (NO), and No.A2411900 (SU). 

\appendix
\section{Energy scale calibratios}
\begin{figure}[h]
\begin{center}
\caption{Example of the fitting of spectra sorted by detector coordinates.
The axis units are the same as those in previous spectral plots.
These are used in section~\ref{cal}.
}
\label{cal-app:sp}
\includegraphics[scale=.35]{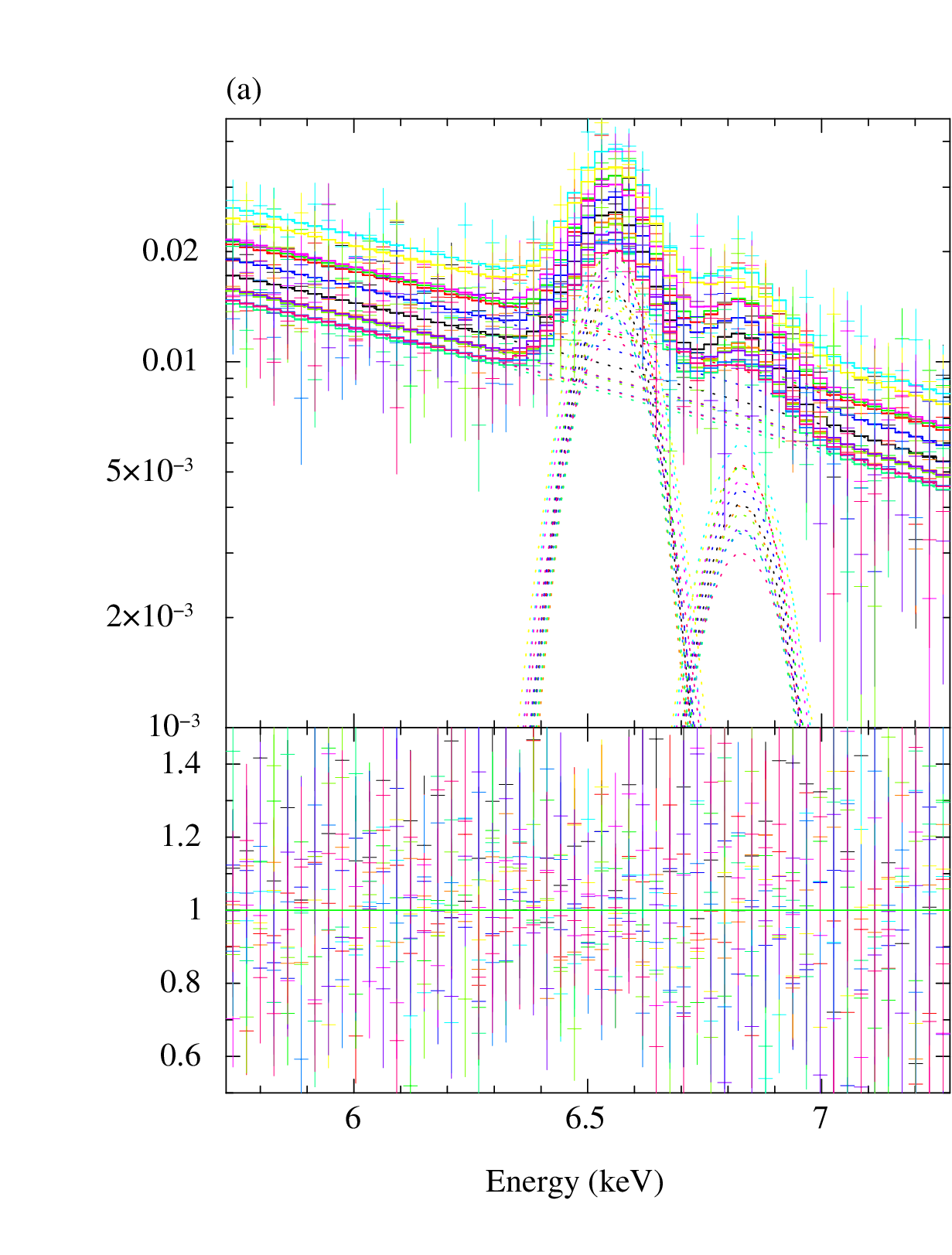}
\includegraphics[scale=.35]{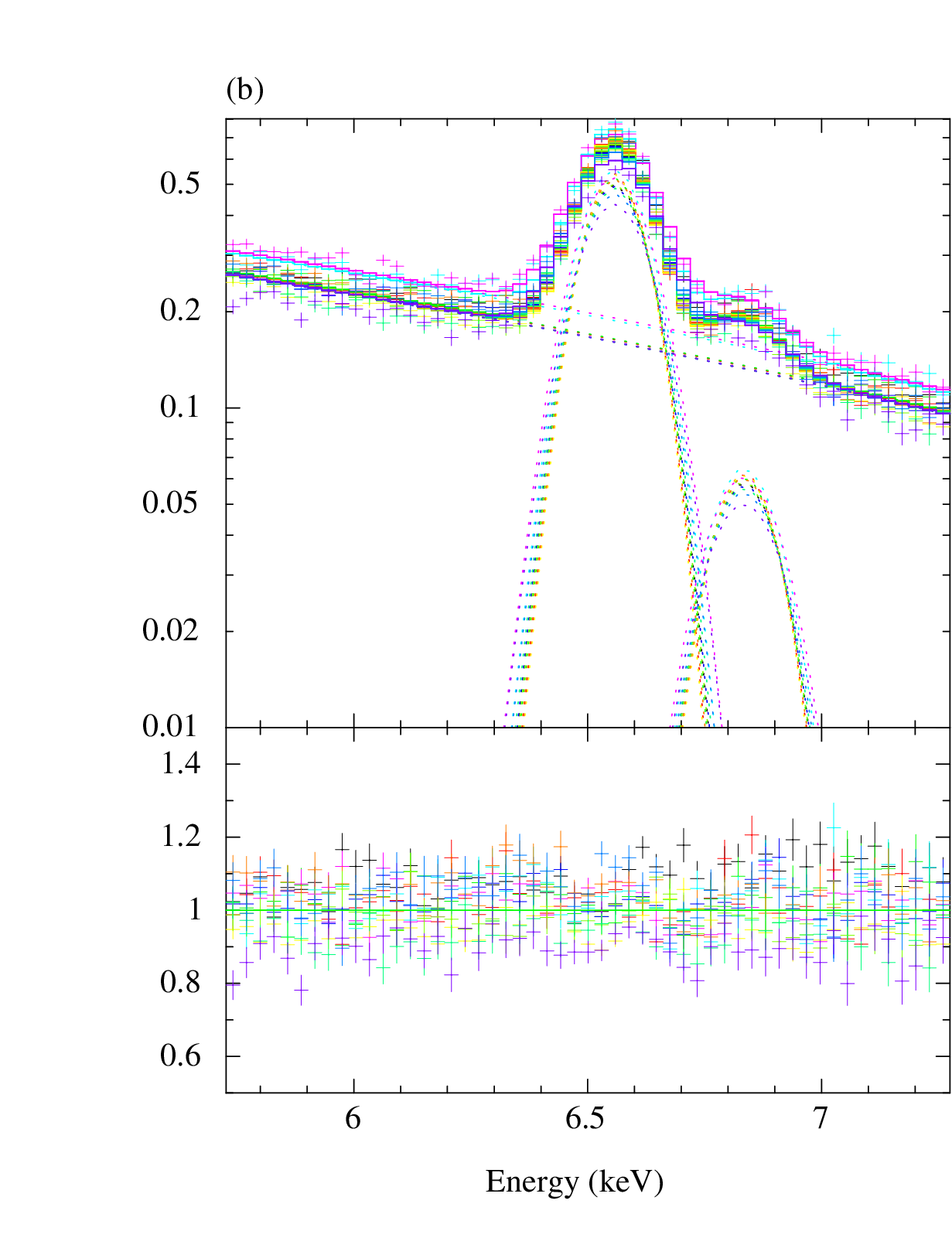}
\end{center}
\end{figure}

\end{document}